\documentclass{acmsmall}

\usepackage{framed}

\usepackage{balance}       
\usepackage{graphics}      
\usepackage[T1]{fontenc}   
\usepackage{txfonts}
\usepackage{mathptmx}
\usepackage[pdflang={en-US},pdftex]{}
\usepackage{color}
\usepackage{booktabs}
\usepackage{textcomp}
\usepackage{enumitem}
\usepackage{cite}
\usepackage{subfigure,xspace,algorithm}
\usepackage[noend]{algorithmic}
\usepackage{color}
\usepackage{amssymb}
\usepackage{epstopdf}

\usepackage{graphicx}

\newcommand{\x}{\mathbf{x}}
\newcommand{\w}{\mathbf{w}}
\newcommand{\f}{\mathbf{f}}

\newcommand{\nop}[1]{}

\newcommand{\ie}{i.e.\xspace}
\newcommand{\eg}{e.g.\xspace}

\begin{document}

\title{Semantic Exploration of Traffic Dynamics}

\title{Semantic Exploration of Traffic Dynamics}
\author{FEI WU
\affil{Pennsylvania State University}
HONGJIAN WANG
\affil{Pennsylvania State University}
ZHENHUI LI
\affil{Pennsylvania State University}}


\begin{abstract}

Given a large collection of urban datasets, how can we find their hidden correlations? For example,  New York City (NYC) provides open access to taxi data from year 2012 to 2015 with about half million taxi trips generated per day. In the meantime, we have a rich set of urban data in NYC including points-of-interest (POIs), geo-tagged tweets, weather, vehicle collisions, etc. Is it possible that these ubiquitous datasets can be used to explain the city traffic? 
Understanding the hidden correlation between  external data and  traffic data would allow us to answer many important questions in urban computing such as: If we observe a high traffic volume at Madison Square Garden (MSG) in NYC, is it because of the regular peak hour or a big event being held at MSG? If a disaster weather such as a hurricane or a snow storm hits the city, how would the traffic be affected?

While existing studies may utilize external datasets for prediction task, they do not explicitly seek for direct explanations from the external datasets. 
In this paper, we present our results in attempts to understand taxi traffic dynamics in NYC from multiple external data sources.  
We use four real-world ubiquitous urban datasets, including POI, weather, geo-tagged tweet, and collision records. 
To address the heterogeneity of ubiquitous urban data, we present carefully-designed feature representations for various datasets. 
Extensive experiments on real data demonstrate the explanatory power on taxi traffic by using external datasets. 
More specifically, our analysis suggests that POIs can well describe the regular traffic patterns. At the same time, geo-tagged tweets can explain irregular traffic caused by big events and weather can explain the abnormal traffic drop. 

\end{abstract}


%
%


\keywords{Urban data, Traffic dynamics, Regression}


\begin{bottomstuff}
This manuscript is an extended version of the paper~\cite{WWL16}.

This work is supported by the National Science Foundation, under
grant  \#1618448 and \#1544455.
Author's addresses: F. Wu, {and} H. Wang, {and} Z. Li, College of Information Sciences and Technology, Pennsylvania State University.
\end{bottomstuff}

\maketitle


\section{Introduction}

Traffic is the pulse of the city that impacts daily life of millions of people. Traffic congestion can make drivers become frustrated and also generate a lot of city noises and vehicle accidents. Therefore, there has been a longstanding strong demand to understand and forecast traffic under different scenarios. An insightful analysis on traffic dynamics could lead to intelligent transportation systems that make the city flow more smooth and make people's life easier.

Modeling traffic dynamics, however, is very difficult as the traffic varies significantly over space and time and it is impacted by many factors simultaneously. To date, various approaches have been proposed to model and to predict traffic~\cite{ZCWY14,CDSC12,OkSt84,XZY07,VDW96,CHMZ11,ARI15,SZY06,XKKL14,CCQ14} with or without considering external context datasets. 
But most of these studies focus on the prediction problem rather than seeking a direct explanation.

\begin{figure}[h!]
\centering
  \includegraphics[width=0.95\columnwidth]{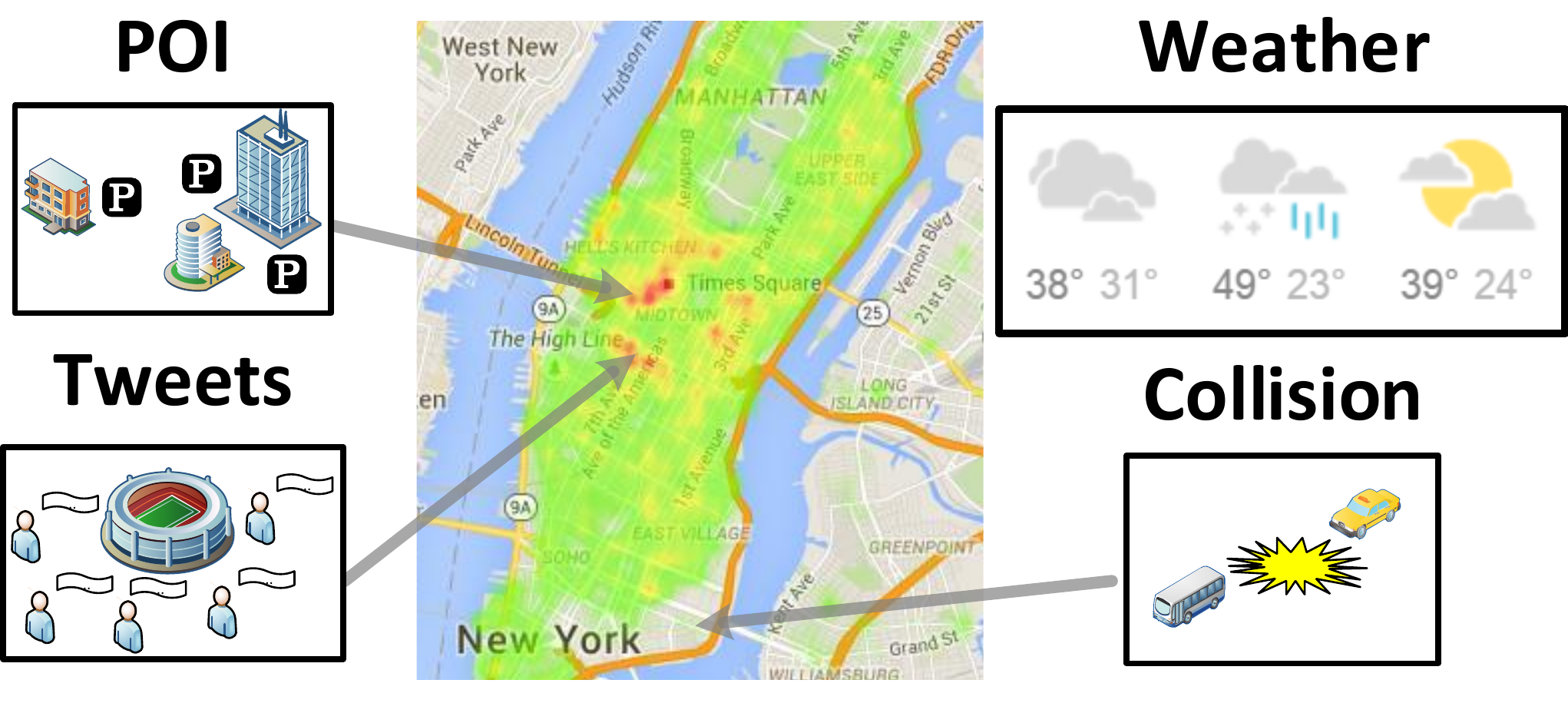}
  \caption{Using ubiquitous urban data to explain traffic.}
  ~\label{fig:intro}
\end{figure}

Motivated by this goal, we propose to study a novel problem: interpreting traffic data using external contextual urban data. Moreover, the big data era has brought us unprecedented urban data, which enables us to take a systematic approach to address this problem. Take New York City for example. The city generates about half million medallion taxi trips; all these data from year 2009 to year 2015 are publicly available on \url{www.nyc.gov} under  the Freedom of Information Law (FOIL). The NYC taxi data was first made public in 2014. It is the first massive public traffic dataset which contains extremely rich information about the urban dynamics in NYC. 

In the meantime, to understand such a large-scale traffic data, several contextual urban data in NYC are being collected from different sources. For example, information about more than 380K points-of-interest (POI) can be collected from FourSquare; people generate about half million geotagged tweets per day in the city; National Centers for Environmental Information provides daily climate information with 28 weather attributes collected from a monitoring tower in Central Park; more than 769K vehicle collisions from 2012 to today are available on NYC open data website (\url{data.cityofnewyork.us}). Our key insight is that all these ubiquitous urban datasets could potentially be valuable signals to explain the traffic dynamics: POIs describe the functions of a region (e.g., a business district always attracts a large volume of morning taxi drop-offs and evening taxi pick-ups; an area with many nightclubs has increased taxi drop-offs at night and pick-ups after midnight). Geo-tagged tweets capture local events (e.g., a popular event will generate peaks in drop-offs before the event and pick-ups after the event). Extreme weather could lead to traffic decline. Vehicle collisions might cause temporary road closures and traffic jams.

While all these urban datasets could potentially explain traffic dynamics, we face several key challenges when modeling the actual correlations:
 \emph{First, all these ubiquitous urban datasets are in different formats and contain different information.} For example, POIs are mostly static and have categorical information such as POI categories as well as numeric information such as number of check-ins. Geo-tagged tweets are timestamped text. Weather information has a lower spatial and temporal granularity and also contains a large portion of missing values.
\emph{Second, the relationship between traffic and the raw value of individual factor is unknown and non-linear.} For example, most fluctuations of weather may not have a big impact on traffic unless it is an extreme weather; popularity of tweets might be associated with people tweeting about a popular global event (e.g., super bowl) and does not necessarily impact traffic. 
Furthermore, some factor (e.g., weather) may have impacts on top of other factors (e.g., POIs). For example, weather could have a damping impact on traffic but will not change the overall shape of the traffic time series, which is largely defined by the regional functions captured by the POIs.

To address these challenges, we study the properties of each dataset and capture their unique impacts by a carefully designed feature representations. We further propose a ridge regression model with polynomial kernel to describe the non-linear and non-additive relation between features. 
We quantitatively evaluate the effects of various datasets on traffic.
Furthermore, qualitative investigations in selected regions lead to many interesting interpretations of the traffic dynamics. 
The analysis suggests that POI data are particularly good in capturing the regular traffic patterns,
while tweets  and weather  are useful under certain occasions. And we do not observe significant impacts of collisions on taxi pick-ups and drop-offs.
We also show that, by using external urban data only, one can infer the traffic condition more accurately compared to methods using historical traffic data only. 

In summary, our paper has three major contributions:
\begin{itemize}[leftmargin=*]
\parskip -0.5ex
\item We study a novel and important problem in urban computing: understanding traffic using ubiquitous urban data. 
\item We investigate ways to design features and build models to capture the correlations between traffic and different types of urban data.
\item We conduct extensive experiments on five large-scale real datasets. The analysis provides insights on how external datasets can be used for interpreting taxi traffic.
\end{itemize}

The rest of the paper is organized as follow. We review the literature in Section~\ref{sec:relatedwork}. Section~\ref{sec:data} describes the taxi dataset. Section~\ref{sec:feature} describes the contextual urban datasets that will be used to explain the taxi data and how we design features based on the properties of these datasets. Section~\ref{sec:method} presents the ridge regression model with polynomial kernel. Section~\ref{sec:exp} presents the empirical evaluations of our method followed by a study of its application in traffic prediction. Section~\ref{sec:discussion} discusses some interesting insights of our discoveries and limitations of the current method. And we conclude our work in Section~\ref{sec:conclusion}. 


\begin{figure*}[t!]
\centering
\subfigure[A normal day (12/10/2012)]{\includegraphics[width=0.28\textwidth]{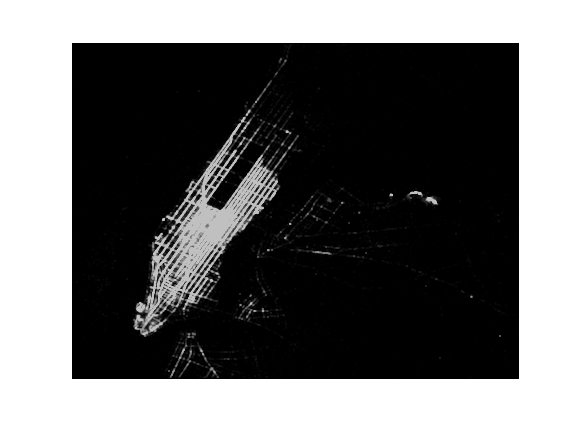}}
\subfigure[Christmas day (12/25/2012)]{\includegraphics[width=0.28\textwidth]{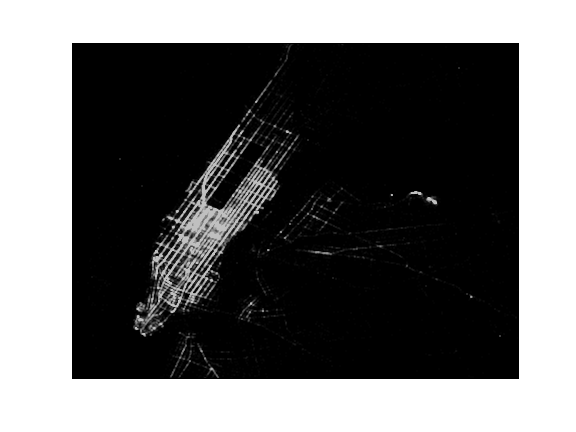}}
\subfigure[Hurricane Sandy day (10/29/2012)]{\includegraphics[width=0.28\textwidth]{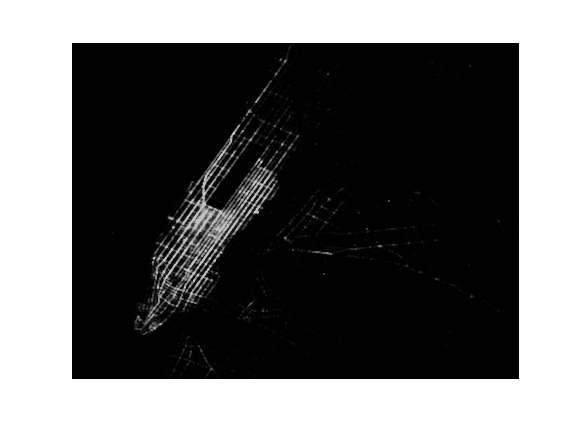}}
\caption{Density map of NYC taxi pick-ups for different days. The difference of these density maps indicates the traffic are highly dynamic over time.}
\label{fig:data}
\end{figure*}

\section{Related Work}
\label{sec:relatedwork}

\textbf{Trajectory prediction of moving objects.} Quite a few previous studies~\cite{MPTG09,Song+13,Song+14,FSSA15,JSZ08,SQBB10,ZhNi12}  have been done on the problem of trajectory prediction. That is, given the historical trajectory traces of individual moving objects, predict the location for a future timestamp. The prediction is based on crowd behavior~\cite{MPTG09,Song+13,Song+14,FSSA15} or routine behaviors~\cite{JSZ08,SQBB10,ZhNi12}. Unlike this line of research on individual movements, our work focus on understanding the aggregated crowd traffic. The latter is very difficulty, more dynamic, computational expensive, and is important for  urban issues such as transportation and public safety.

\textbf{Traffic  prediction.} Traffic  prediction has been extensively studied in transportation research area. Representative forecasting models include neural network models~\cite{CDSC12}, Kalman filters~\cite{OkSt84,XZY07}, autoregressive integrated moving average (ARIMA) models~\cite{VDW96,CHMZ11,ARI15},  Bayesian network approach~\cite{SZY06,XKKL14}, and Markov Random Fields~\cite{CCQ14}. 

Our problem and method are different from these studies in the following aspect. First, our problem is to model the correlation between traffic and external urban datasets instead of forecasting future traffic based on historical traffic data. Second, all these studies only consider one dataset -- traffic data. The external factors that may have a huge impact on traffic are ignored. Our problem takes into consideration several large-scale urban datasets such as POI, geo-tagged tweets, weather, and vehicle collisions and study their impacts on traffic. Third, since our objective is not to predict future traffic based on history, the existing models cannot be applied. We need a different model to describe the complicated correlations with various impacting factors.


\textbf{Computing with heterogeneous urban data.} In recent years, urban computing~\cite{ZCWY14} has gain an increasing popularity due to availability of large-scale urban data. These studies include using different urban datasets such as POI, taxi, bike rental, or noise complaint to profile city functions~\cite{YZX12}, detect traffic anomalies~\cite{ZZY15}, predict air quality~\cite{ZLH13,Zhen+15}, gas consumption~\cite{Shan+14}, and location recommendation~\cite{Lian+15}. 
Our work is under the same theme of urban computing in the context of large-scale heterogeneous data.
However, we distinctly put our emphasizes on understanding the traffic patterns from external sources.


\section{Traffic Data}
\label{sec:data}

\begin{figure}
\centering
  \includegraphics[width=0.6\columnwidth]{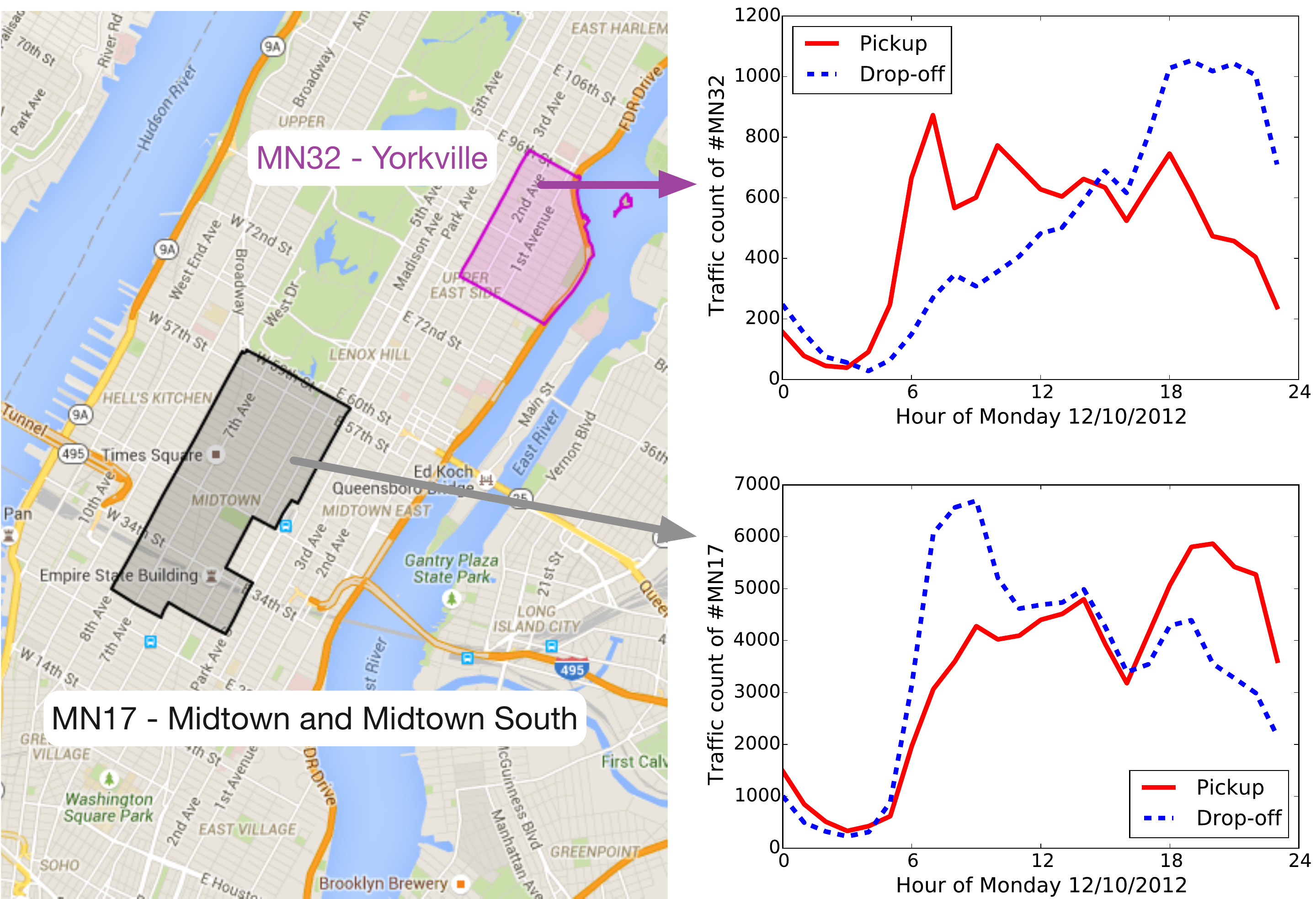}
  \caption{Traffic patterns are significantly different for different neighborhoods. Yorkville is a residential neighborhood with a heavy pick-up traffic in the morning and a significant number of drop-offs in the evening. The Midtown shows a reverse traffic pattern because the neighborhood is one of the largest business district in Manhattan. }
  \label{fig:data-loc}
\end{figure}

A large-scale New York City taxi dataset has been made public online~\cite{nyctaxi}. 
The dataset contains all trips completed in yellow and green taxis from 2009/01/01 to 2015/12/31. Each trip contains information about pick-up location and time, drop-off location and time, trip distance, fare amount, etc. 
We use the subset of trips from 2012/10/1 to 2012/12/31, which has $\bf  28,759,878$ trips. On average, we have $463,869$ trips per day. We pick this time period, as it aligns with the date of collection of other external context datasets.  
From the raw data, we make the following observations.

\textbf{Traffic density changes over time.}
Figure~\ref{fig:data} shows the distribution of GPS points of the pick-up and drop-off locations on different occasions. 
We can see that the traffic on holidays (\eg, Christmas as shown in Figure~\ref{fig:data}(b)) and under extreme weather condition (\eg, Hurricane Sandy as shown in Figure~\ref{fig:data}(c)) are significantly less compared with traffic in a normal day (as shown in Figure~\ref{fig:data}(a)).

\textbf{Different traffic patterns at different locations.}
Figure~\ref{fig:data-loc} shows that traffic patterns vary at different neighborhoods, as different neighborhoods have different functions. 
For Yorkville neighborhood (which is mainly consisted of residential area), we see a peak for pick-ups in the morning and a peak for drop-offs in the evening on weekday.
For Midtown and Midtown South neighborhood, which mainly serves business purposes, a sharp peak for drop-offs appears in the morning whereas the peak for pick-ups occurs in the evening. 

The observations indicates the potential correlations exist between traffic and contexts, \eg, weather and function of the region.
In the following section, we further investigate several urban datasets and study their correlations between with the traffic.

\begin{figure*}[t!]
\centering
\subfigure[Nightlife]{\includegraphics[width=0.15\textwidth]{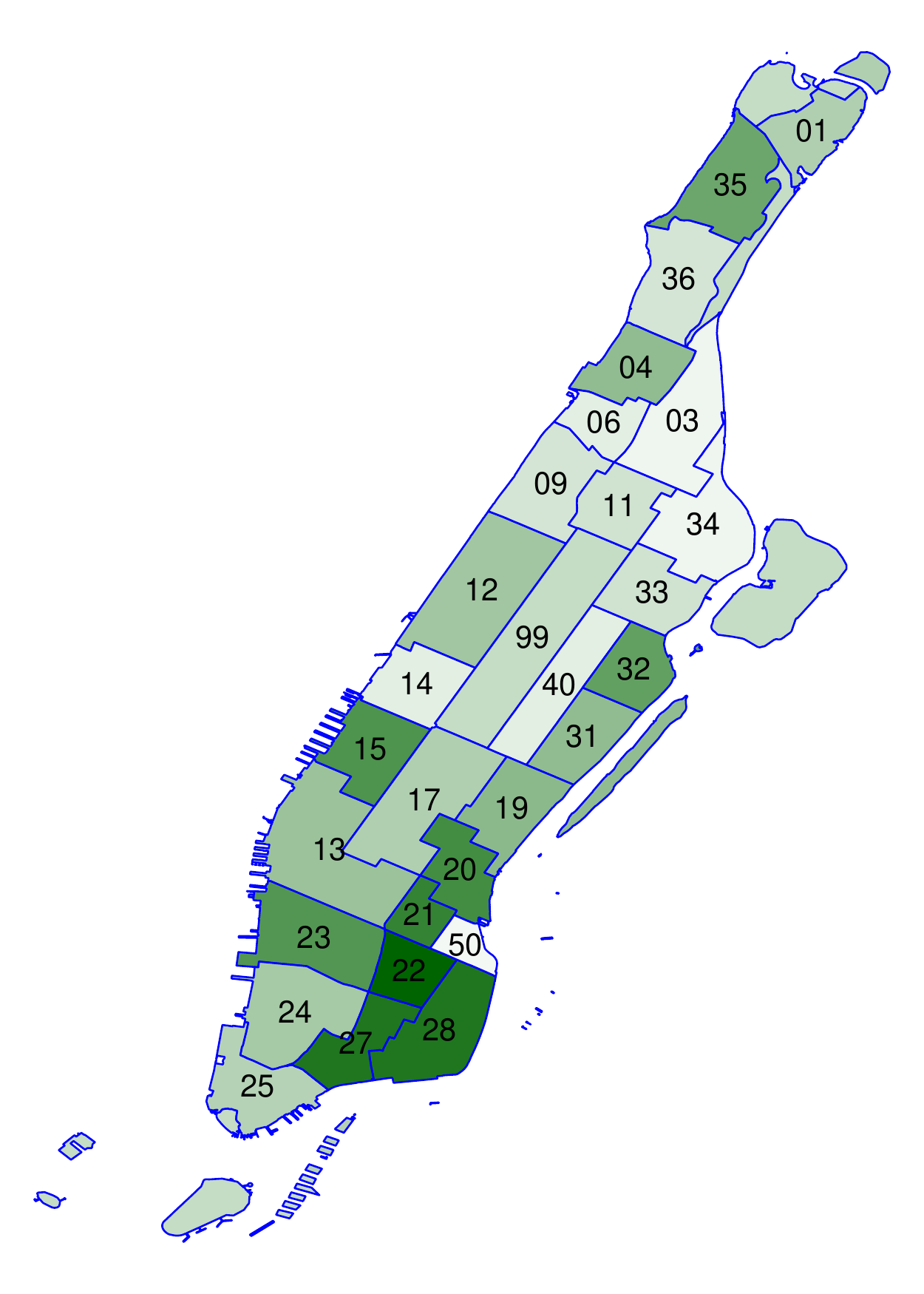}}
\subfigure[Shops]{\includegraphics[width=0.15\textwidth]{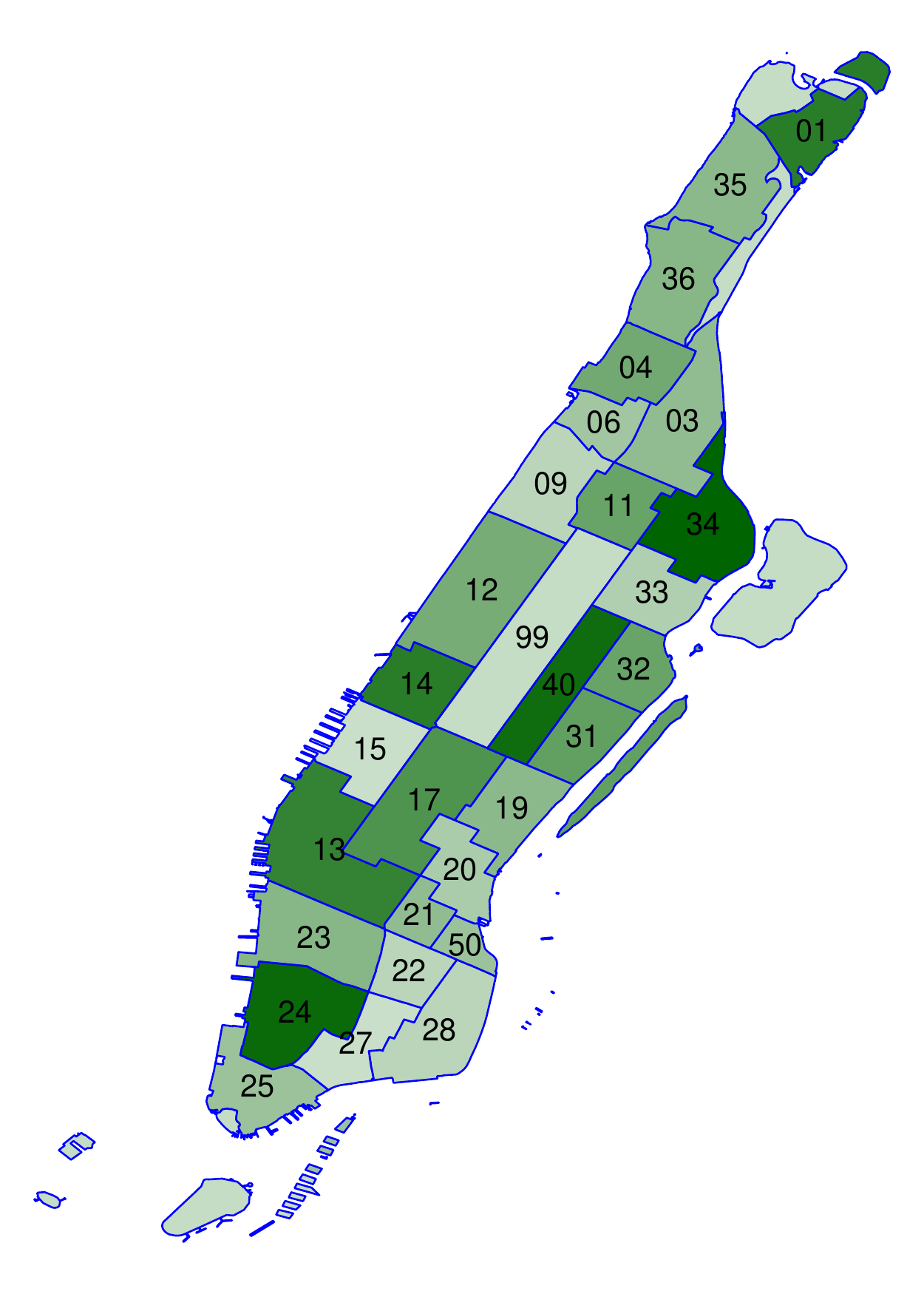}}
\subfigure[Arts \& Entertainment]{\includegraphics[width=0.15\textwidth]{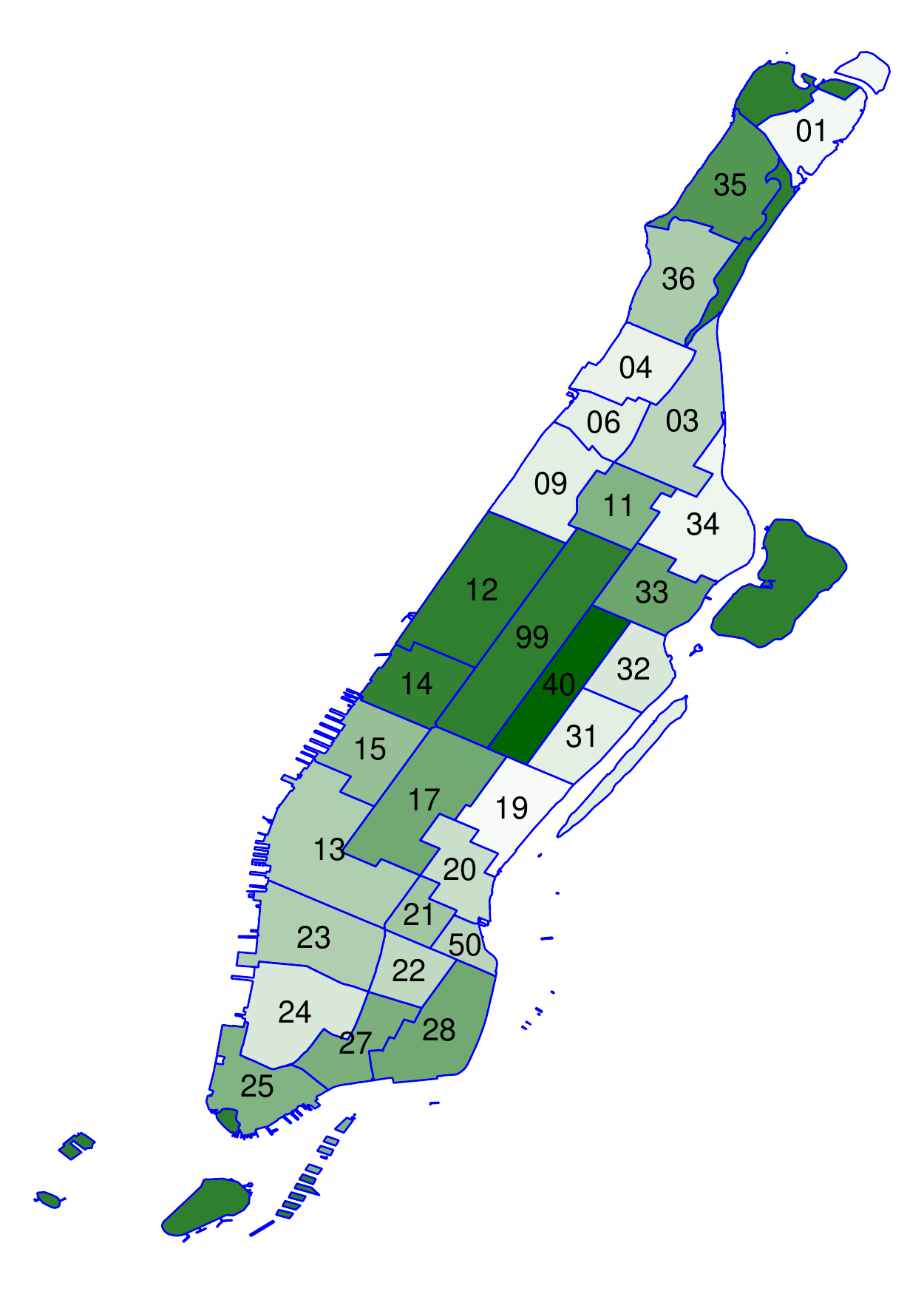}}
\subfigure[College \& Education]{\includegraphics[width=0.15\textwidth]{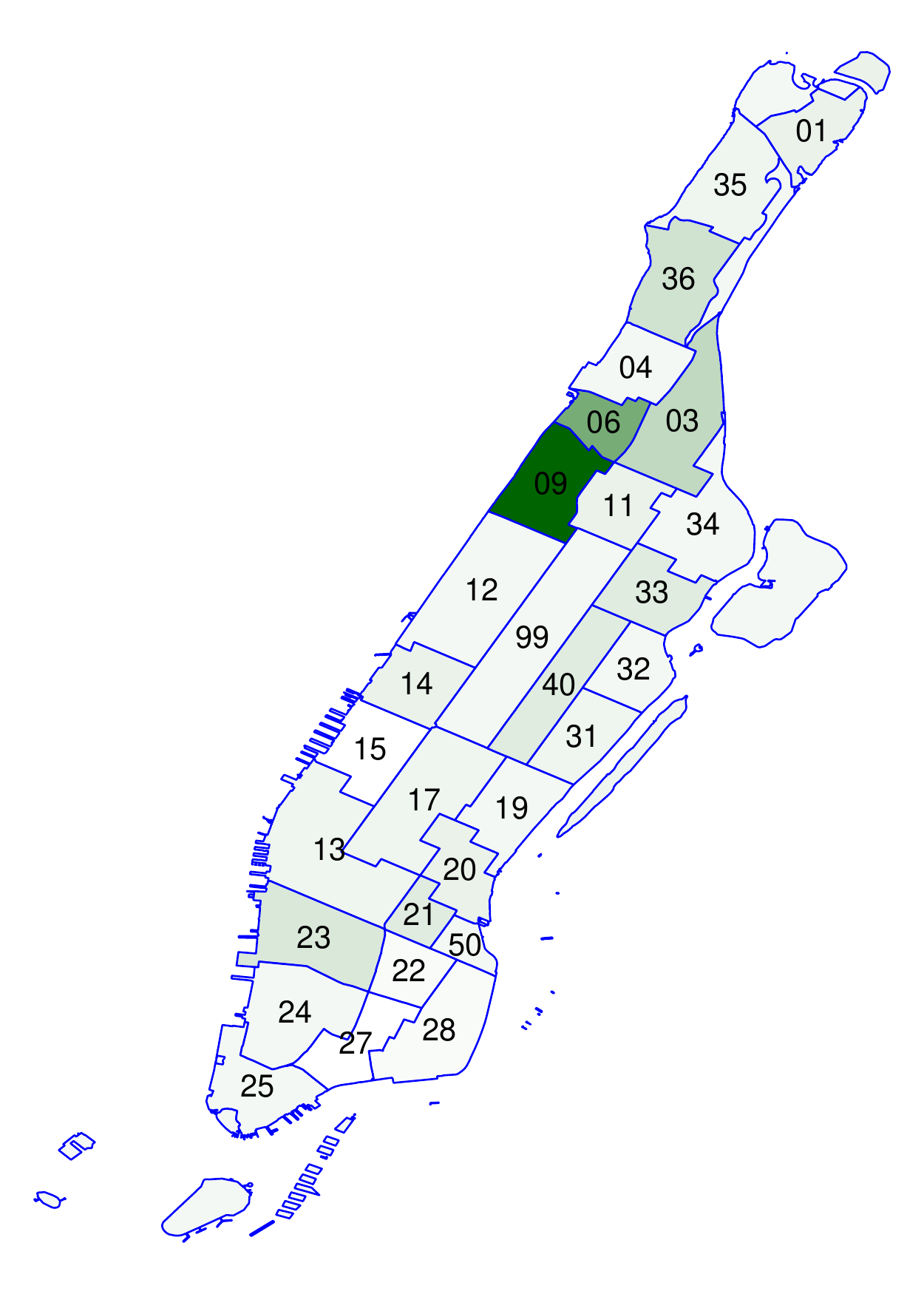}}
\subfigure[Outdoors \& Recreation]{\includegraphics[width=0.15\textwidth]{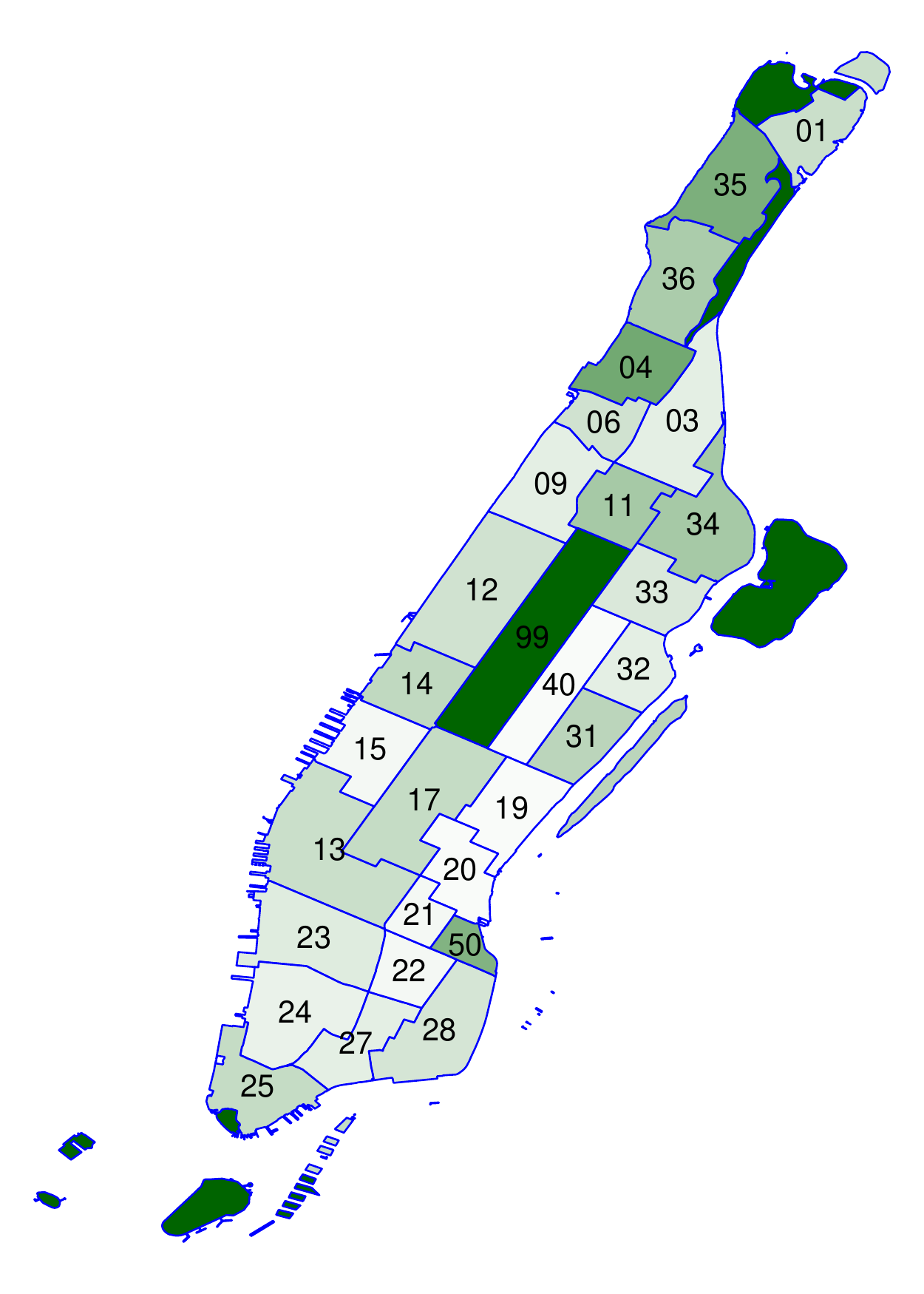}}\\
\subfigure[Professional]{\includegraphics[width=0.15\textwidth]{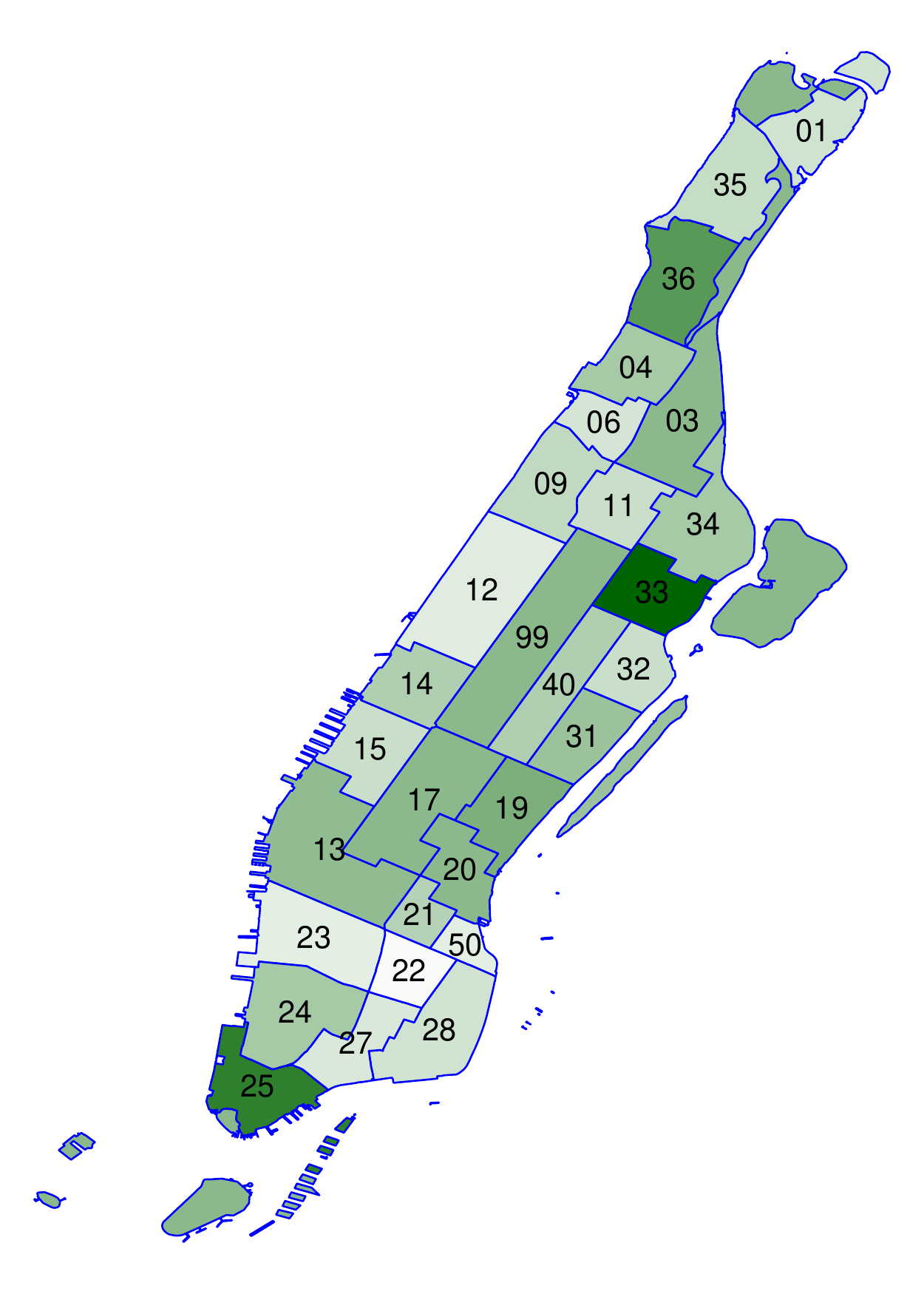}}
\subfigure[Travel]{\includegraphics[width=0.15\textwidth]{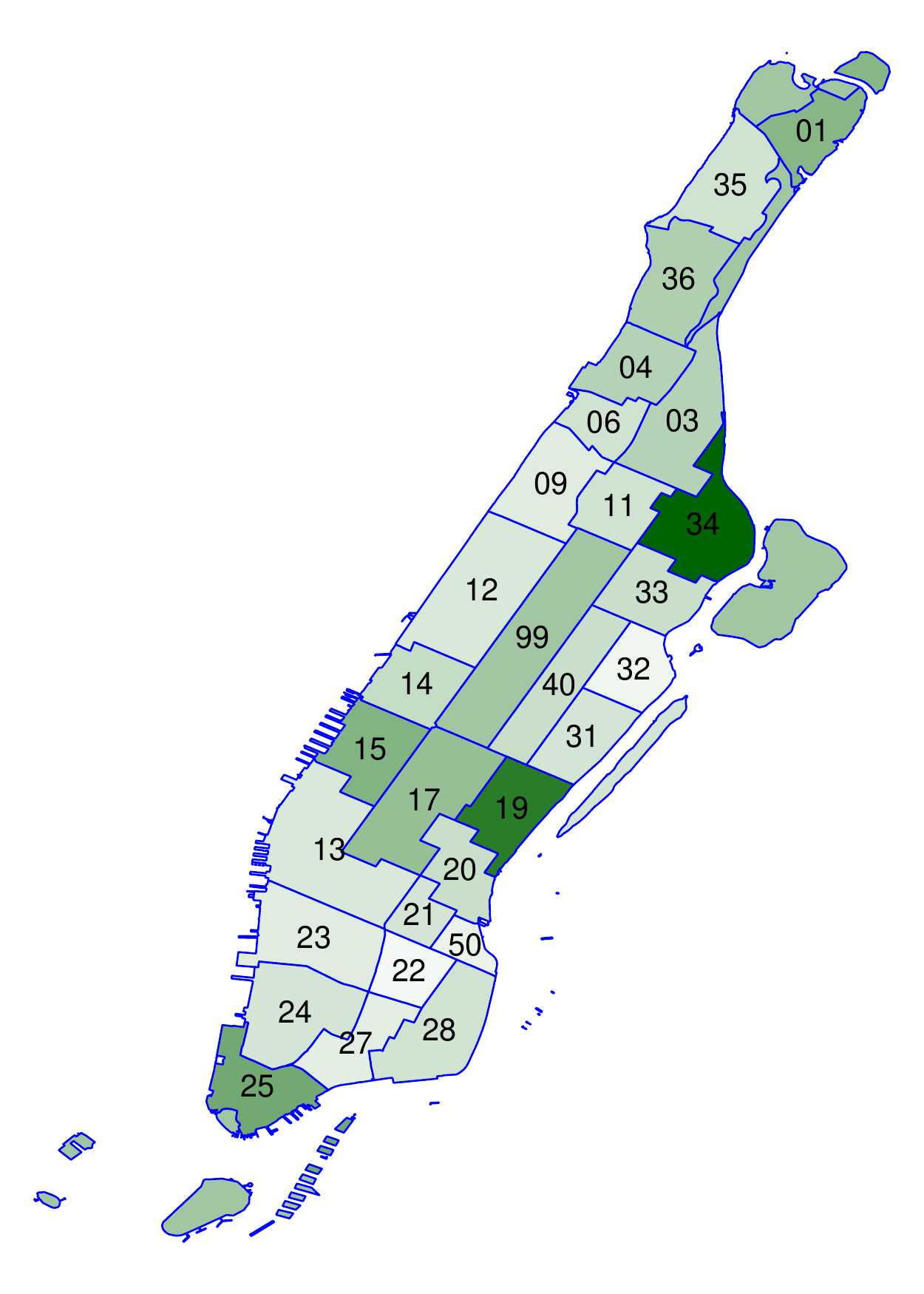}}
\subfigure[Residence]{\includegraphics[width=0.15\textwidth]{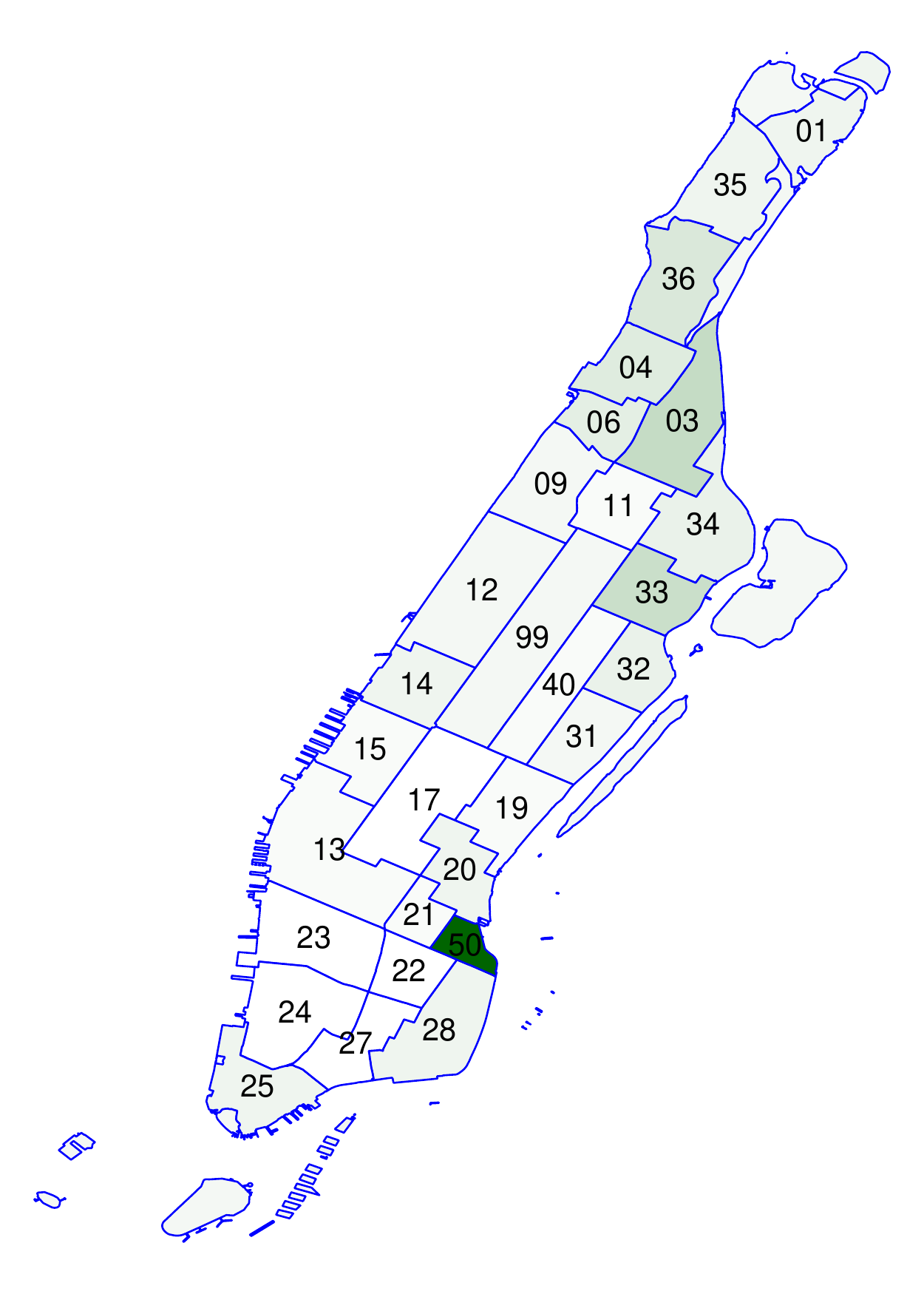}}
\subfigure[Food]{\includegraphics[width=0.15\textwidth]{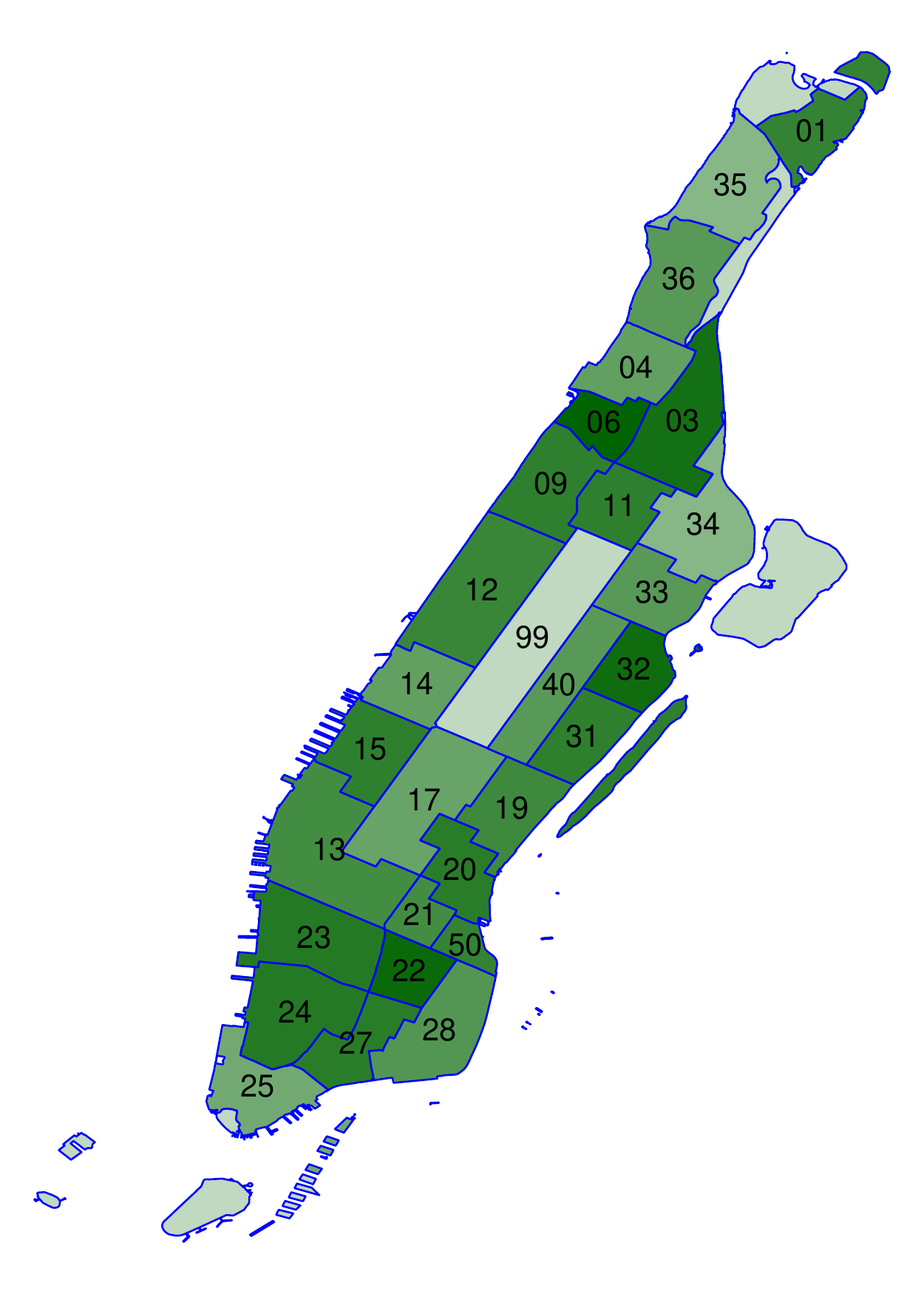}}
\subfigure[Event]{\includegraphics[width=0.15\textwidth]{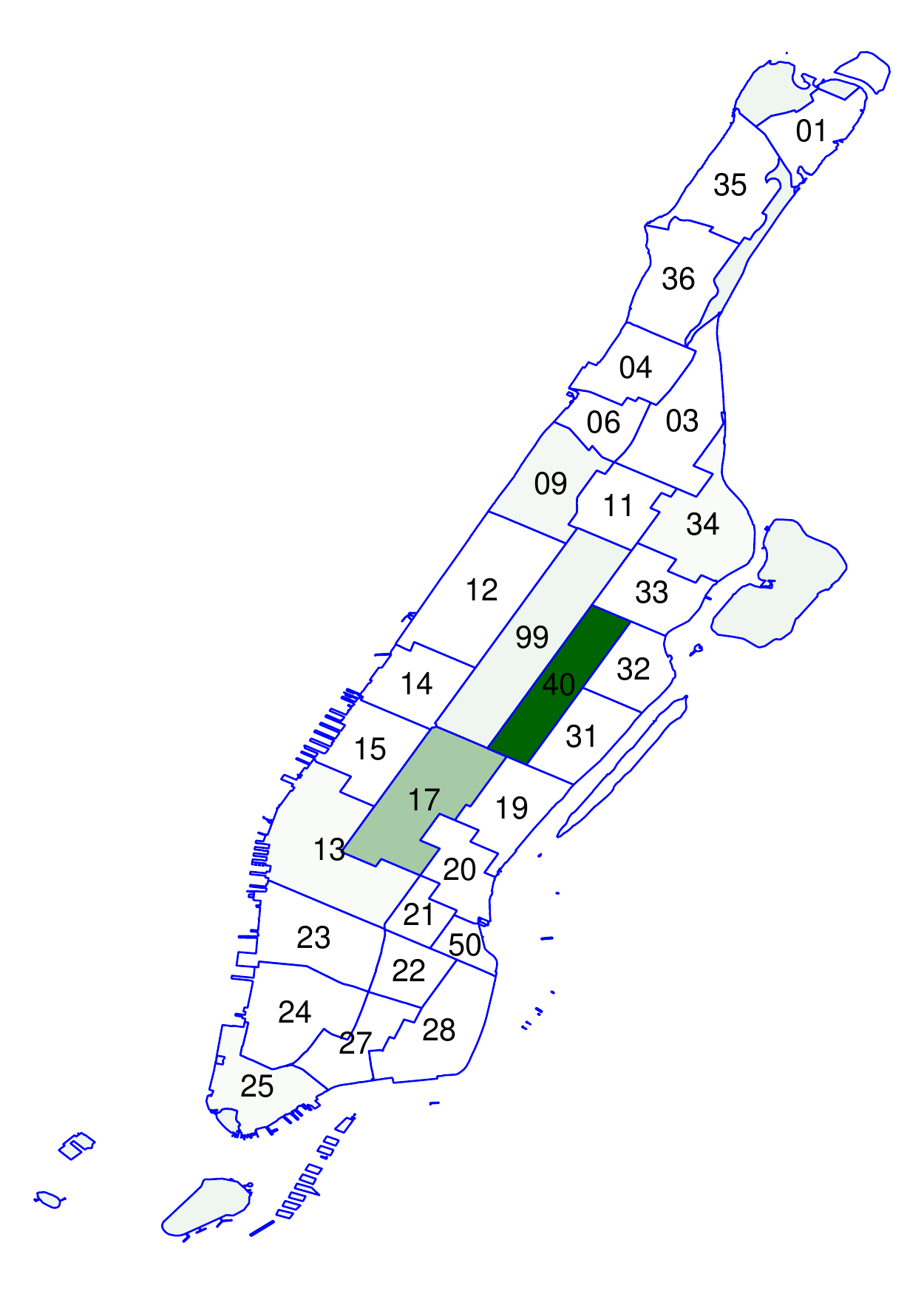}}
\caption{POI category distributions in Manhattan of NYC.}
\label{fig:poi-coef}
\end{figure*}

\section{Feature Design from Explanatory Urban Data }
\label{sec:feature}
\subsection{Point-of-Interest (POI) for Regional Functions}
POI database provides us with rich information about the venues, which potentially correlates with the functionality of each location. 
Therefore, we gather a POI dataset from FourSquare~\cite{foursquare}.
The FourSquare API provides us venue information such as venue name, category, number of check-ins, and number of unique visitors. 
We focus on the major category information to characterize the neighborhood functions. 
There are 10 first-level categories in total: {food, residence, travel, arts \& entertainment, outdoors \& recreation, college \& education, nightlife, professional, shops, and event.}

We query the FourSquare API to gather venues that are checked in  by at  least  one  user.  
In total, we get information about 380,380 venues for New York City. 
Figure~\ref{fig:poi-coef} shows the number of unique users checked in to each category of the POIs for all neighborhoods in Manhattan area.  
The darker the color is, the higher the number. 
The color map suggests the degree of functions for each neighborhood with respect to a specific category. 
From Figure~\ref{fig:poi-coef}(a), we can see that majority of the users go to South East corner for nightlife. In Figure~\ref{fig:poi-coef}(c), POIs near Central Park are more related to arts $\&$ entertainment.

\begin{figure}[t!]
\centering
  \includegraphics[width=0.65\columnwidth]{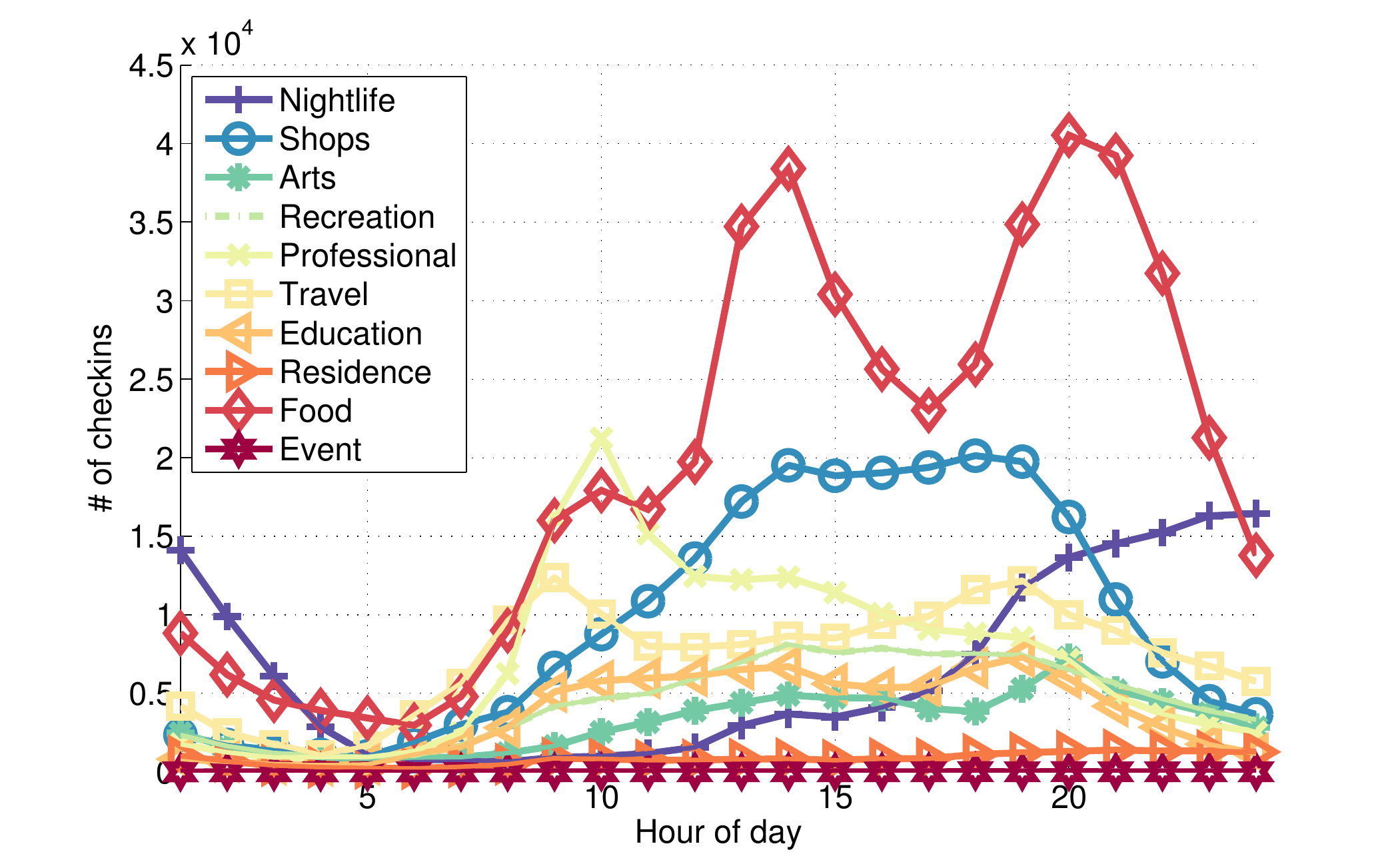}
  \caption{POI popularity over 24 hours in a day.}
  ~\label{fig:poi-time-series}
\end{figure}

While the POI dataset enables us to model the overall functions of each region, 
it is important to consider the popularity over time for those categories. 
The time-varying popularity indicates the time span during which a region may be of interest to people.
Therefore, we obtain dataset of FourSquare check-in posts from Twitter following the strategy introduced in~\cite{NSMP11}. 
We crawl geo-tagged tweets over the same time period, i.e., from Oct, 2012 to Dec, 2012.
The check-in tweets start with \textit{``I'm at"} followed by a venue name.
A check-in tweet is matched with a POI if: (i) the geo-tag of the tweet is within 100 meters of the POI location, and (ii) similarity between POI names is large than 0.8, where the similarity is defined as the length of the longest common subsequence divided by the length of the longer sequence.  
As a result, we obtain $1,598,617$ check-ins. 
Later, the check-ins are aggregated by the POI category and by the hour of the day. 
Figure~\ref{fig:poi-time-series} shows the time-varying popularity (i.e., number of check-ins) of each category in each hour of the day.
We can see that the popularity of food related venues peak at noon and early night times, while nightlife venues (\eg, bars) are more popular during night and have little attention during day time. 

Given the two datasets, we define POI features for each region.
Formally, let $\mathbb{C}$, $\mathbb{D}$, and $\mathbb{T}$ denote the set of all categories, the set of all spatial grid cells, and the set of all timestamps, respectively. We consider a set of $n$ POIs on the map: $\mathbb{P} = \{p_1, p_2, \ldots, p_n\}$. Each POI $p_i$ is represented as a tuple $(c, l, z)$, where $p_i.c$ is the category of this POI, $p_i.l$ is its geographic location, and $p_i.z$ is the overall popularity of this POI measured by the total check-ins from FourSquare (this data is directly obtained from FourSquare API).
We calculate the POI feature value for category $c\in \mathbb{C}$ in grid cell $d\in \mathbb{D}$ at time $t\in \mathbb{T}$ as:
\begin{equation}
\mathbf{f}_{POI}(c, d, t) = \sum_{i: p_i.l \in d \land p_i.c = c} p_i.z \times g(c,P(t)),
\end{equation}
where $P(t)$ is the relative time (\ie, hour of the day) of $t$, $g(c,P(t))$ is the temporal popularity of the category $c$ at relative time of $t$, \ie, shown in Figure~\ref{fig:poi-time-series}. 
We adapt this hourly popularity definition for simplicity.
There are other options for defining the temporal popularity, such as further differentiating between work day and non-work day, or aggregate the values at a different time granularity.

\subsection{Geo-Tagged Tweets for Local Events}
The POI features may help us capture the regular traffic at each location. 
To capture irregularity in traffic, we seek to extract event occurrences using geo-tagged tweets.
Again we use the geo-tagged tweets dataset we collected in NYC around the same time period (from November, 2012 to Dec, 2012). 
Each geo-tagged tweet is of the form  $\langle timestamp,userid,latitude,longitude,content\rangle$.

We count number number of unique users posting tweets that containing hashtags.
Intuitively, local events should be identifiable via peaks in the space and time.
Formally, we define the tweet feature value for a grid cell $d\in \mathbb{D}$  at time $t\in \mathbb{T}$ as:
\begin{equation}
\mathbf{f}_{tweet} (d, t) = c(d, t),
\end{equation}
where $c(d, t)$ is the count of distinct users post a tweet at grid cell $d$ at time $t$.
We count the number of users instead of the posts to alleviate the problem of spammers.
The number of users provides us a signal of population density at the location at the given time.

\subsection{Weather for Disasters}
Intuitively, the traffic should correlate with weather.
Furthermore, extreme weather conditions such as hurricane and snow storm could significantly impact traffic. 
To capture the impact of weather, we use the daily weather dataset in USA from National Centers for Environmental Information~\cite{noaa}. 
There is one climate monitoring tower in Manhattan that situated in Central Park. 
The data contains 28 weather attributes for each day. 
We use the highest 2-mins wind speed, highest 5-seconds wind speed, precipitation, and snow fall information, as they are indicators for extreme weather in NYC.

Extreme weather conditions may have a lasting effect on traffic, as it may take days for recovering from damage.
Therefore, we model the effect of extreme weather conditions by a power-law decaying function.
Formally, we define the impact of an extreme weather event $e$ at time $t$ (after the event) on its corresponding weather feature $a$ as:
\begin{equation}
\mathbf{f}_{e}(a,t) =  max\{c(a,t_e)- \lambda(t-t_e)^{\alpha},0\}, t\in [t_e,\infty)
\end{equation}
where $\alpha$ and $\lambda$ are positive values controlling the decay, $t_e$ is when event $e$ happens, and $c(a,t_e)$ is the value of weather feature (corresponding to the extreme weather event) at time $t_e$. $\mathbf{f}_{e}(a,t)$ is 0 for $t<t_e$.
When $\alpha > 1$ ($\alpha < 1$) the function decays slower (faster) at the beginning for a small time lag and gradually decays faster (slower). 
We let $\alpha>1$ to capture the lasting effect of severe weather conditions. 
\begin{figure}[h!]
\centering
  \includegraphics[width=0.45\columnwidth]{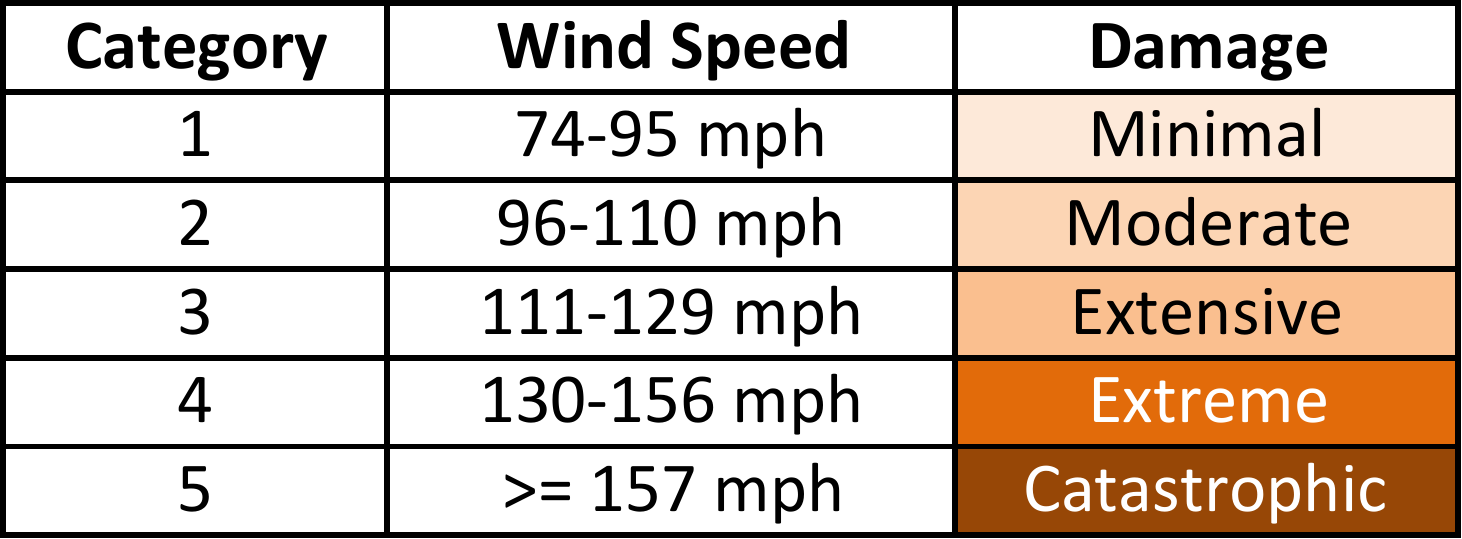}
  \caption{Saffir-Simpson hurricane wind scale.}
  ~\label{fig:hurricane-category}
\end{figure}
In practice, the values of $\alpha$, and $\lambda$ depend on the severity of the weather conditions, the magnitudes of observed values, and the types of weather condition. 
We determine whether a severe weather event happens according to Saffir-Simpson hurricane wind scale~\cite{nhc} on wind speed and
according to Northeast Snowfall Impact Scale~\cite{snow} on snowfall. 
In our experiment, we tune the two parameters to make the effect of extreme values last for  three days.
An example of wind scale categories and severity is shown in Figure~\ref{fig:hurricane-category}.
We also consider a severe weather event happens if the feature value is larger than $\mu+3\sigma$ ($\mu$ and $\sigma$ are mean and standard deviation of the feature values).

Finally, we can define weather features at time $t$ as:
\begin{equation}
\mathbf{f}_{weather}(a,t)= \sum_{e\in E_a}\mathbf{f}_{e}(a,t),
\end{equation}
where $E_a$ is set of all extreme weather events related to feature $a$.

\subsection{Vehicle Collisions for Traffic Jam}

Vehicle collisions could potentially cause traffic controls on certain blocks and thus impact local traffic. Details of Motor Vehicle Collisions in New York City provided by the Police Department (NYPD) and can be accessed from NYC Open Data~\cite{collision}. This data is constantly being updated. From July 1st, 2012 to March 14th, 2016, there are about 769K collisions. 
To align with other datasets, we use data from the same period, \ie, from Oct, 2012 to Dec 2012.

Let $\mathbb{R}$ denote the set of all collisions in our study. Each collision record $r_i\in \mathbb{R}$ is represented as a tuple $(t, l, s)$, where $r_i.t$ the time of the collision, $r_i.l$ is the latitude and longitude of the collision, and $r_i.s$ is the number of people injured or killed in this collision. We construct our feature for collisions on day $t$ in grid cell $d$ using the number of collisions weighted by the severity of the collisions:
\begin{equation}
\mathbf{f}_{collision}(d, t) = \sum_{i: r_i.t = t \land r_i.l in d} (r_i.s+1).
\end{equation}

\section{Model}
\label{sec:method}
We now present our approach to model the correlation between traffic and urban features. For each location grid $d$, we denote the number of pick-up (or drop-off) at time $t$ as $y_t$. The feature space is a combination of all the features:
\[
 \x_t = [\f_{POI}(:,d, t),\; \f_{tweet}(d, t),\; \f_{weather}(:,t),\; \f_{collision}(d,t)].
 \] 
Here, $\f_{POI}$ is a 10-dimensional feature because there are 10 categories for the POIs, $\f_{tweet}$ and $\f_{collision}$ are one-dimensional features, and $\mathbf{f}_{weather}$ is $k$-dimensional feature where $k$ depends on the number of weather attributes used. In our experiment setting, we use wind speed, precipitation, and snow fall.
Given the data $\{y_t, \x_t\}_{t=1}^N$ for a location grid $d$, our goal is to fit a regression model $y = f(\x)$ for that grid. 

Linear regression is arguably the most frequently used regression model in practice. It asserts that the response is a linear function of the inputs:
\begin{equation}
y_t = \w^T \x_t + w_0 = \sum_{j=1}^{D} w_j x_j^t + w_0,
\end{equation}
where $w_j$ and $x_j^t$ are the $j$-th entry in the vectors $\w$ and $\x_t$, respectively, $D$ is the dimension of the feature vector, and $w_0$ is a bias term.

To fit the model with training data, we need to find the parameter $\w$ that minimizes the error:
\begin{equation}
J(\w) = \frac{1}{N} \sum_{t=1}^{N} \left(y_t - (w_0 + \w^T \x_t)\right)^2.
\end{equation}

A common technique to avoid overfitting is to control the values of weights $\w$ on the features, where we add an $L_2$-regularization to the objective function.
In addition, the features may have multiplicative interactions among them.
For example, in a bad weather day, the total traffic could decrease, but the relative traffic pattern over time (e.g., peak hours vs. non-peak hours) remains similar, which can be described by the temporal popularity of POIs. In this case, the traffic is determined as a combined effect of weather feature and POI feature.
To avoid overfitting as well as consider the multiplicative effects, 
we use Kernel ridged regression (KRR) as our regression model.


\section{Empirical Evaluation}
\label{sec:exp}
In this section, we demonstrate the effectiveness of our approach with large-scale real world urban datasets. We first present the result of quantitative evaluation on our entire dataset. 
Later, we further conduct qualitative investigation on several regions of interest in hope of understanding how traffic data correlates with external datasets.

\subsection{Quantitative Evaluation}
\textbf{Setting.}
We divide Manhattan into 500 meter $\times$ 500 meter grids and build a model for each grid to predict hourly number of taxi pick-ups and drop-offs.
There are 319 grids in the middle and lower area of Manhattan (\ie, South of $86 th$ street). 
Our experiments are conducted on these grids where the data from different sources are less sparse.  
Similar grid partition strategy has also been used in previous study~\cite{ZhWa15}.
We use the first two weeks as training (from 10/01/2012 to 10/15/2012) to fit the model and use the next week for testing (from 10/16/2012 to 10/23/2012). 
The average hourly drop-off is 53 (pick-up is 56) for all grids and the standard deviation for hourly drop-off is 38 (pick-up is 41).
The traffic also varies for different grids with the highest hourly drop-off being 926 (pick-up being 828) and lowest hourly drop-off (and pick-up) being 0.

We use mean-square error (MSE) and mean relative error (MRE) to evaluate testing error (on testing data) and coefficient of determination ($R^2$) to evaluate fitness of the model (on training data). 
Given a set of $M$ test samples $\{y_i, \x_i\}_{i=1}^M$, the rooted mean-square error is defined as: $RMSE = \sqrt{\frac{1}{M}\sum_i(y_i-\hat{y_i})^2}$, where $y_i$ is the ground truth value, and $\hat{y_i}$ is estimated value. 
The mean-relative error is defined as $MSE = \frac{1}{M}\sum_i|\frac{(y_i-\hat{y_i})}{y_i}|$.
The coefficient of determination $R^2$ is defined as $1- \frac{Var(Y-\hat{Y})}{Var(Y)} \in (-\infty, 1]$. 
Note that the higher the $R^2$ value is, the better the data fit a model. 

\nop{
\begin{table}[t]
\centering
\begin{tabular}{cc}
\begin{tabular}{|c||c|c||c|}
\hline 
Drop-off     & RMSE   & MRE & $R^2$\\  \hline \hline
Linear regression & $24.1$ & $0.52$ & $0.62$ \\ \hline
Ridge regression & $24.1$  & $0.52$ & $0.62$ \\ \hline
Kernel Ridge regression& $24.1$  & $\mathbf{0.5}$  & $\mathbf{0.64}$ \\ \hline   
\end{tabular} 
&
\begin{tabular}{|c||c|c||c|}
\hline
Pick-up    & RMSE   & MRE & $R^2$\\  \hline \hline
Linear regression & $26.4$ & $0.50$ & $0.57$ \\ \hline
Ridge regression & $26.4$  & $0.50$ & $0.57$ \\ \hline
Kernel Ridge regression& $\mathbf{26.1}$  & $\mathbf{0.48}$  & $\mathbf{0.59}$ \\ \hline     
\end{tabular}

\end{tabular}
\caption{Model comparison. Rooted mean-square error (RMSE) Mean-relative error (MRE) and coefficient of determination ($R^2$) are reported.}
\label{table:model}
\end{table}
}

\begin{table}[t]
\tbl{Feature effectiveness. P: POIs, T: geo-tagged tweets, W: weather, C: collision. RMSE, MRE and $R^2$ values are reported for different combination of features. \label{table:feature}}{
\centering
\begin{tabular}{|c||c|c|c|c|c|} 
\hline
	 & all feature groups   & $T+W+C$  & $P+T+C$ & $P+W+C$ & $P+W+T$ \\ \hline 
$R^2$& $0.64$ & $0.07	$ & $0.63$ & $0.62$ & $0.61$ \\ \hline
RMSE  & $24.1$   & $50.9$ & $24.1$ & $23.7$ & $24.1$ \\ \hline
MRE  & $0.5$  & $1.51$ & $0.5$ & $0.5$ & $0.5$ \\ \hline
\end{tabular} }
\end{table}

\textbf{Feature Effectiveness.}
To study feature importance, we use leave-one-out strategy, that is, testing the model performance by excluding one feature from the feature set.
Kernel ridge regression is the model used in this experiment.

Table~\ref{table:feature} summaries the performance with different features (we only report the result for drop-off and the result of pick-up is similar). 
Without using POI features, RMSE is nearly two times larger than the model with POI features included, and MRE is three times larger. Similarly,  $R^2$ is $0.06$ when excluding POI features, while  $R^2$ is always higher than $0.6$ when POI features are used. The result suggests that POI features correlate the best with the taxi traffic data in general. 
At the same time, there is no significant improvement  by including weather data, collision data, and tweet data. This is because traffic follows the routine behavior for most of the time, which is captured by POI features. Other features can describe abnormal scenarios (e.g., events and extreme weather), but such scenarios rarely happen or only happen in some small regions (e.g., convention center). When evaluating overall performance, these features become less important.  However, these features are important in some cases at some specific locations. In the following sections, we qualitatively investigate several regions of interest to understand the correlation between traffic and these features. 

\subsection{Qualitative analysis}

\begin{figure}[t!]
\centering
  \includegraphics[width=0.45\textwidth]{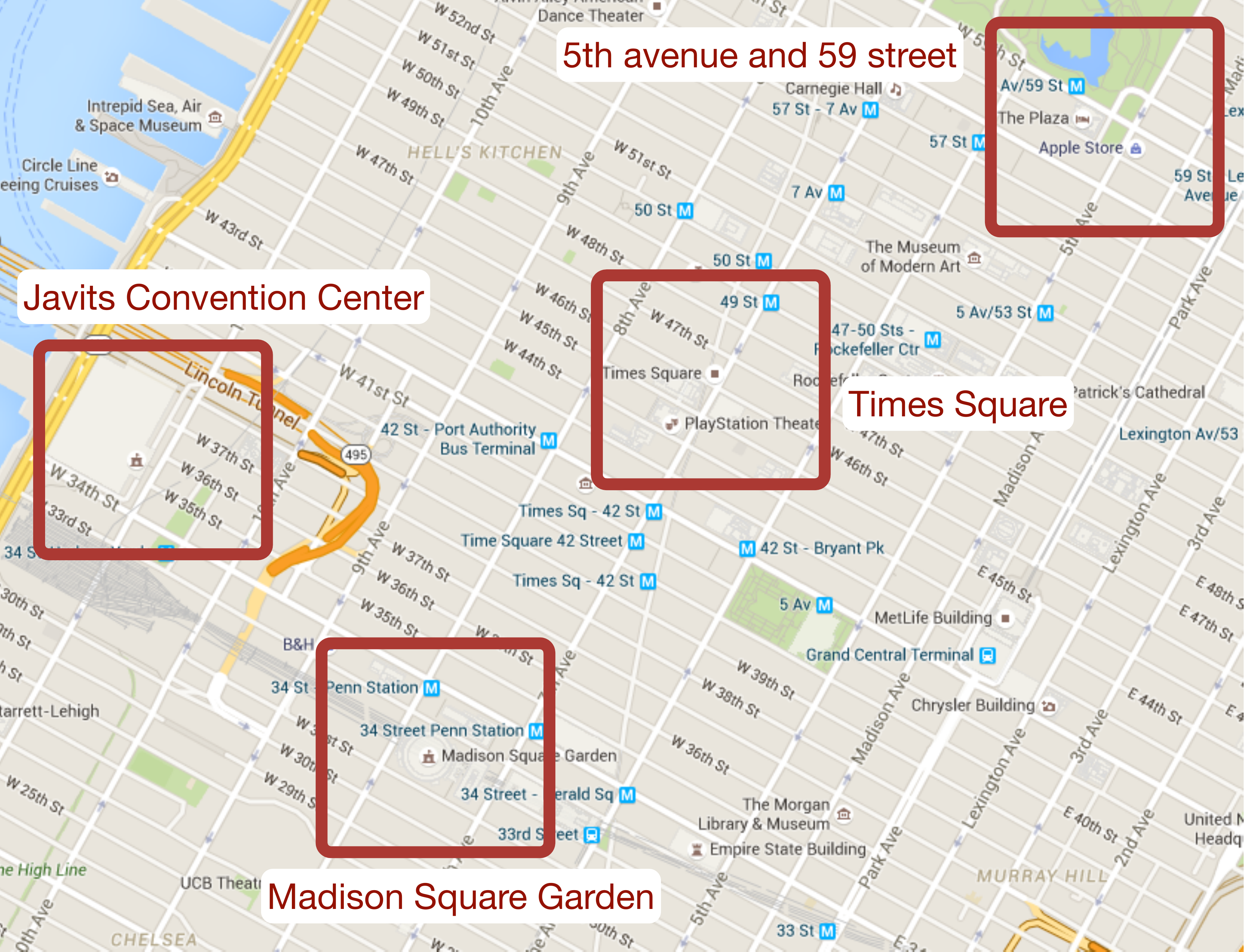}
  \caption{Selected regions of interest for investigation.}
  ~\label{fig:3locs}
\end{figure}

\begin{figure*}[t!]
\centering
\subfigure[Madison Square Garden. Traffic explained by POI. More details in Figure~\ref{fig:poi-explain}(a).]{\includegraphics[width=1\textwidth]{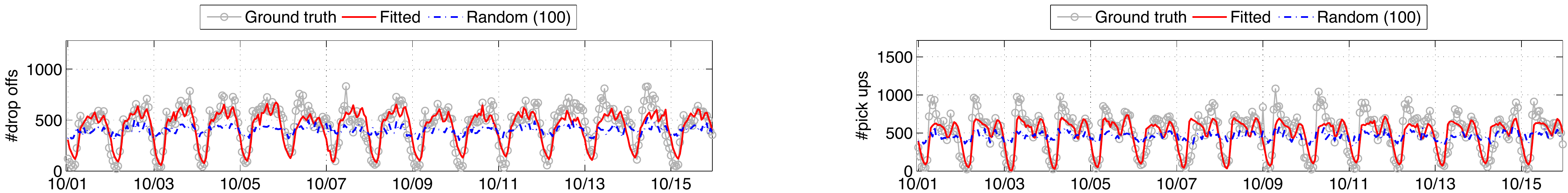}} 
\subfigure[Times Square. Traffic explained by POI. More details in Figure~\ref{fig:poi-explain}(b).]{\includegraphics[width=1\textwidth]{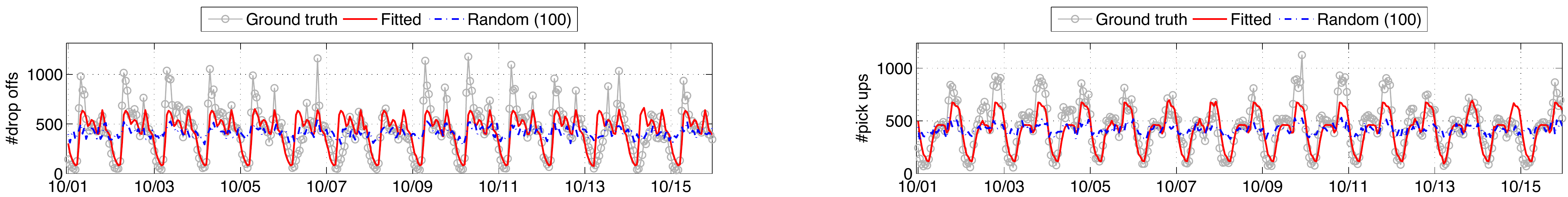}} 
\subfigure[Fifth Avenue. Traffic explained by POI. More details in Figure~\ref{fig:poi-explain}(c).]{\includegraphics[width=1\textwidth]{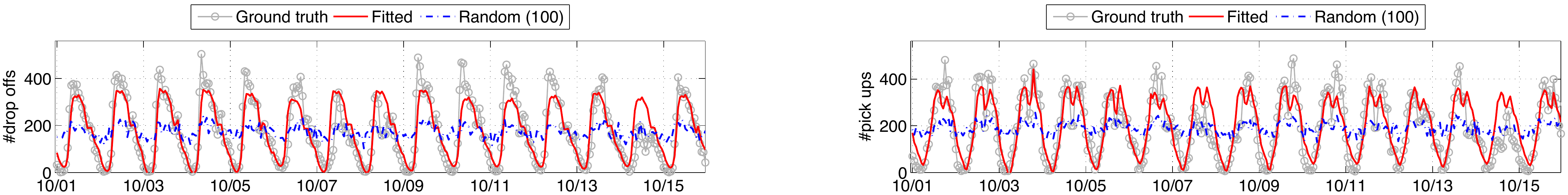}} 
\subfigure[Javits Convention Center. Traffic explained by POI and geo-tagged tweets. More details in  Figure~\ref{fig:nycc} and Figure~\ref{fig:jacob-traffic}.]{\includegraphics[width=1\textwidth]{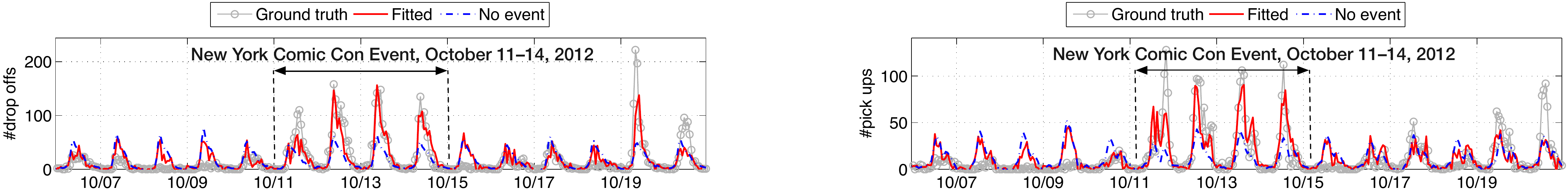}} 
\subfigure[Madison Square Garden. Traffic explained by POI and weather. More details in Figure~\ref{fig:weather}.]{\includegraphics[width=1\textwidth]{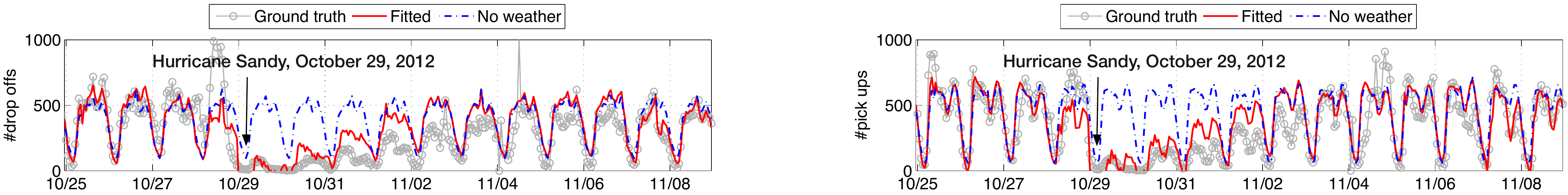}} 
\caption{Fitting results of traffic data using other urban datasets.}
\label{fig:cases}
\end{figure*}

We manually select four regions for further inspection.
We aim to pick a set of representative areas that cover different functions of a city, \ie, a tourist sight with entertainment related venues (Times Square), a big arena and a transportation center (Madison Square Garden, which sits on top of Penn Station), a central business area with many shops (5th avenue), and a large convention center (Jacob K. Javits Convention Center). 
Figure~\ref{fig:3locs} shows the locations of the picked regions on a map. 
Each region is covered by an approximately 500 meter X 500 meter grid. 
We design our analysis with two questions in mind:
\begin{itemize}
\item Q1 (Fitness): Can we construct the taxi traffic patterns by using external urban data?
\item Q2 (Interpretation): Which features are useful and under what circumstances they are useful? 
\end{itemize}
To answer these two questions, we look at the fitting result of our model on all areas, as we want to know whether the external urban data can be used to explain the taxi traffic patterns. 
Using fitting result for explanatory purposes has also been seen in previous studies~\cite{Mats+14,Mats+15} 
Figure~\ref{fig:cases} shows the fitting results for these four regions. 

\begin{figure*}[t]
\begin{tabular}{ccc}
\includegraphics[width= 0.3\textwidth]{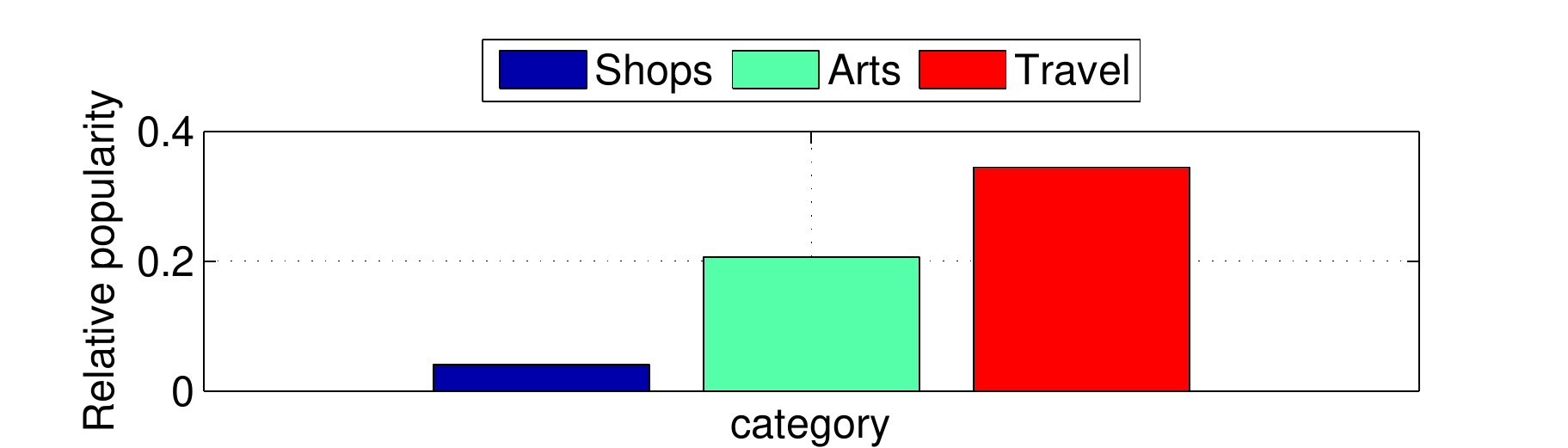} &\includegraphics[width= 0.3\textwidth]{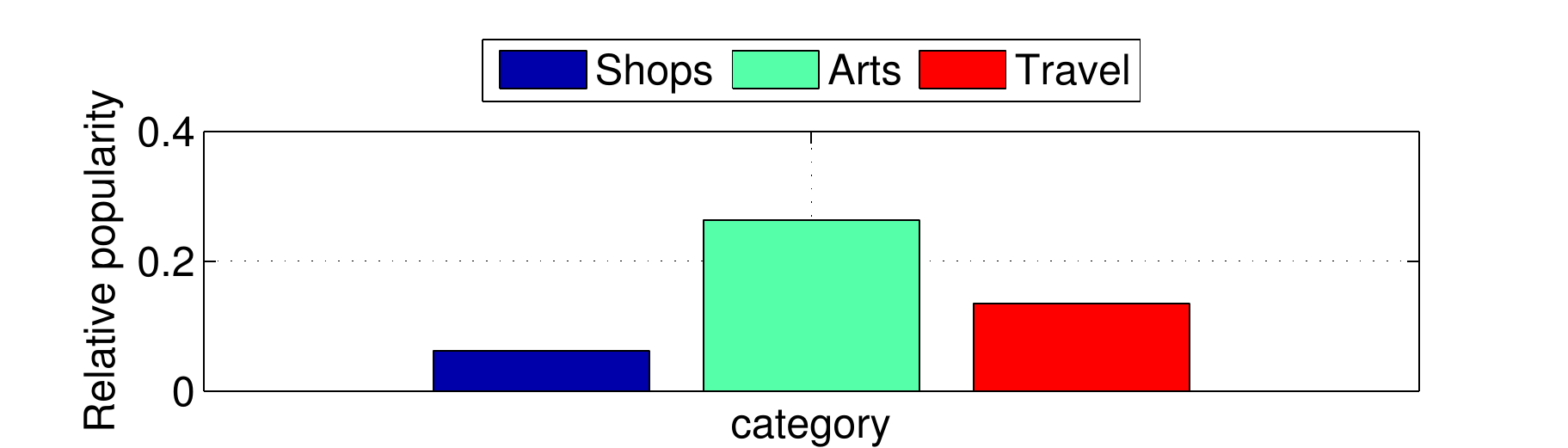}&\includegraphics[width= 0.3\textwidth]{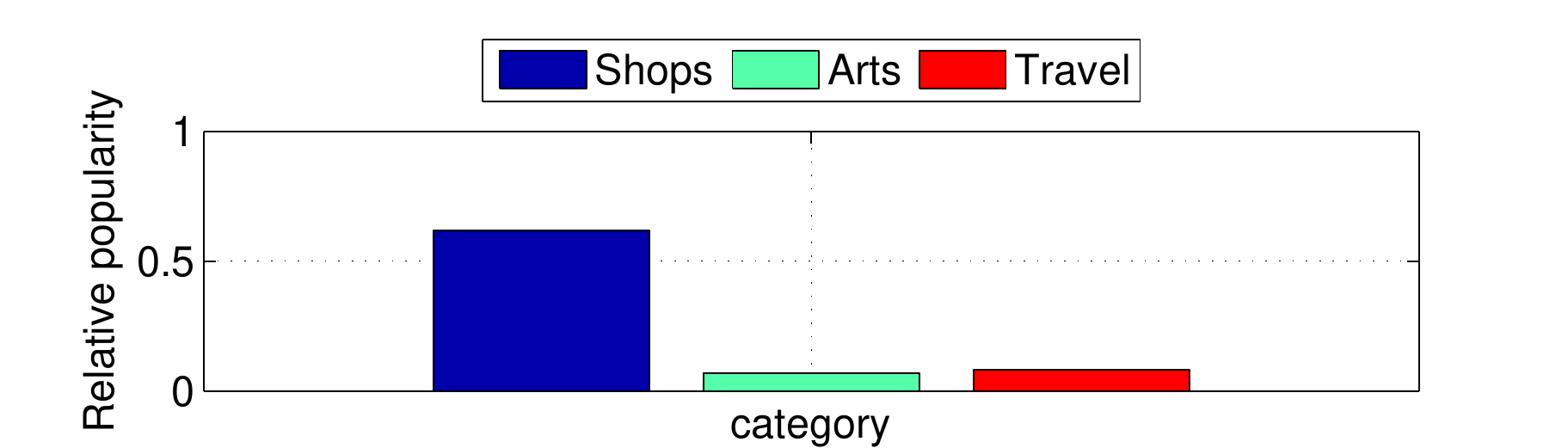}\\
\includegraphics[width= 0.3\textwidth]{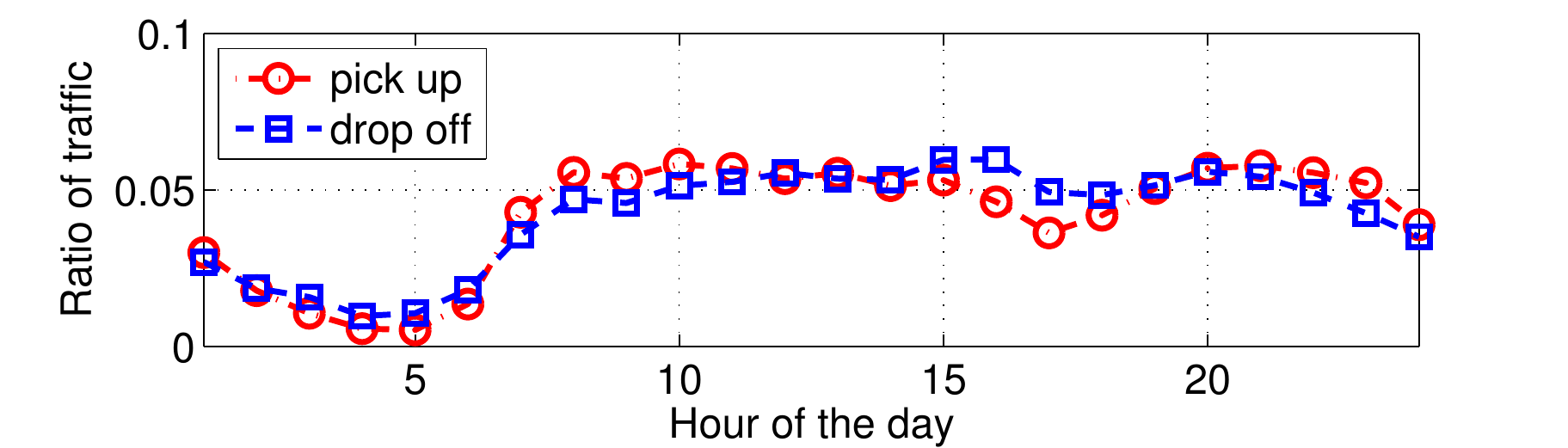} &\includegraphics[width= 0.3\textwidth]{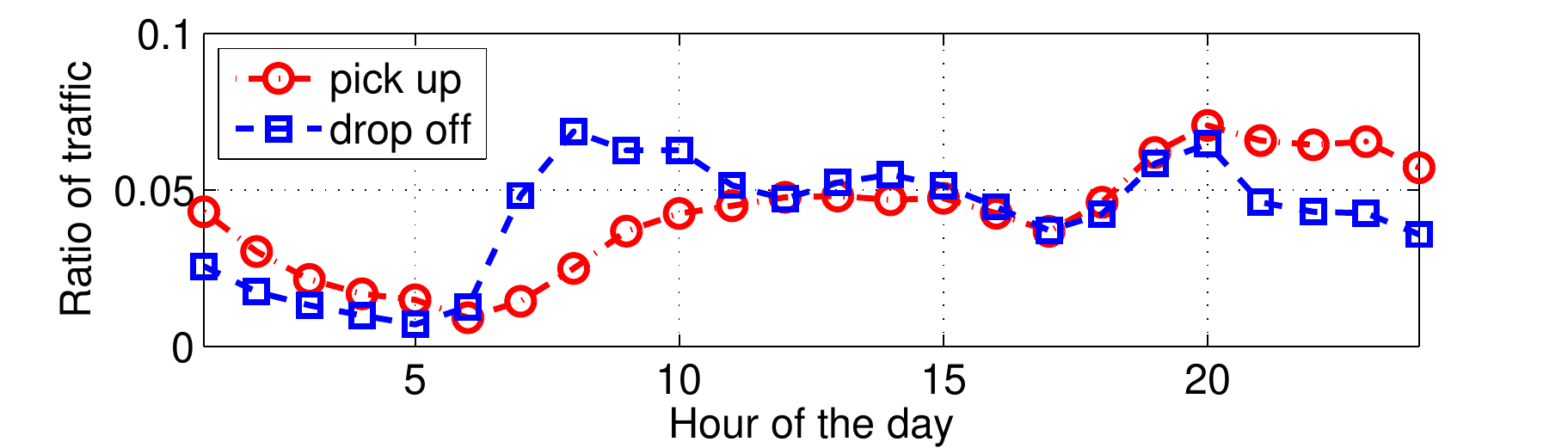}&\includegraphics[width= 0.3\textwidth]{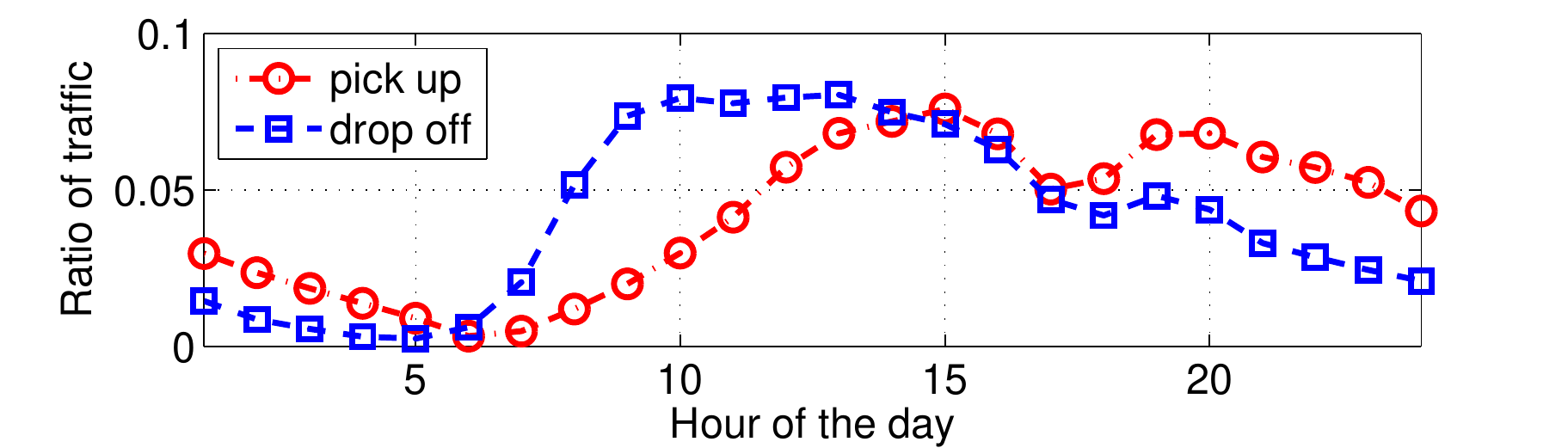}\\
(a) Madison Square Garden & (b) Times Square & (c) Fifth Avenue
\end{tabular}
\caption{POI distribution and traffic pattern for three regions. Madison Square Garden (the same location for Penn Station) has a constantly high traffic volume. Times Square has a high drop-off and pick-up at 8 p.m. because of Broadway shows at night. Fifth avenue has a high drop-off during the daytime because people come to shop and most shops close at 8 p.m. }
\label{fig:poi-explain}
\end{figure*}
\subsubsection{Traffic w.r.t. POIs.} 
Figure~\ref{fig:cases}(a), (b), and (c) show the fitting results using our POI features for drop-offs and pick-ups at Madison Square Garden, Times Square, and 5th Avenue, respectively. In addition, we show the POI distribution (on discriminative categories) of these three regions in Figure~\ref{fig:poi-explain}.
As one can see, in Madison Square Garden, categories ``art'' and ``travel'' dominate the distribution because Madison Square Garden hosts a lot of events and beneath that is Penn Station, one of the biggest transit center in NYC. We observe high traffic volume constantly from 7am to midnight because people transit through there all the time.
For Times Square, category ``art'' is the biggest category because there are many theaters on Broadway. This explains why we see peaks for drop-offs and pick-ups around 8pm. 
Finally, 5th Avenue has a higher proportion of POIs in the ``Shop'' category. Therefore, we observe many drop-offs in the morning and pick-ups at 7pm (most shops close around 8pm -- 9pm). 

To further verify the external data indeed provide useful information, we generate 100-dimensional features with random values and use the same kernerlized ridge regression model to fit the traffic data (shown as blue dashed line in Figure~\ref{fig:cases}(a), (b), and (c)). Obviously, these random features, although with a much higher dimension, cannot fit the traffic data well. This demonstrates the effectiveness of our POI features in explaining the traffic. 
The result also aligns with our quantitative experiment where POI features show strong correlation overall the regions.

\subsubsection{Traffic w.r.t. Geo-tagged Tweets.}
Geo-tagged tweets may help explain unexpected traffic patterns where local events are dominant source of traffic.
Although overall geo-tagged tweets do not correlate with traffic, we found cases where traffic patterns can be attributed to ongoing events at a large-event venue, \ie, Javits center.
As shown in Figure~\ref{fig:cases}(d), we look at traffic data at Javits center during the period
from 10/11/2012 to 10/13/2012, where there is large event, \ie, NYC Comic Con.
It is clear that the considering geo-tagged tweets significantly improve the fitness of our model  for event days. 

\begin{figure}[t!]
\centering
  \includegraphics[width=0.8\columnwidth]{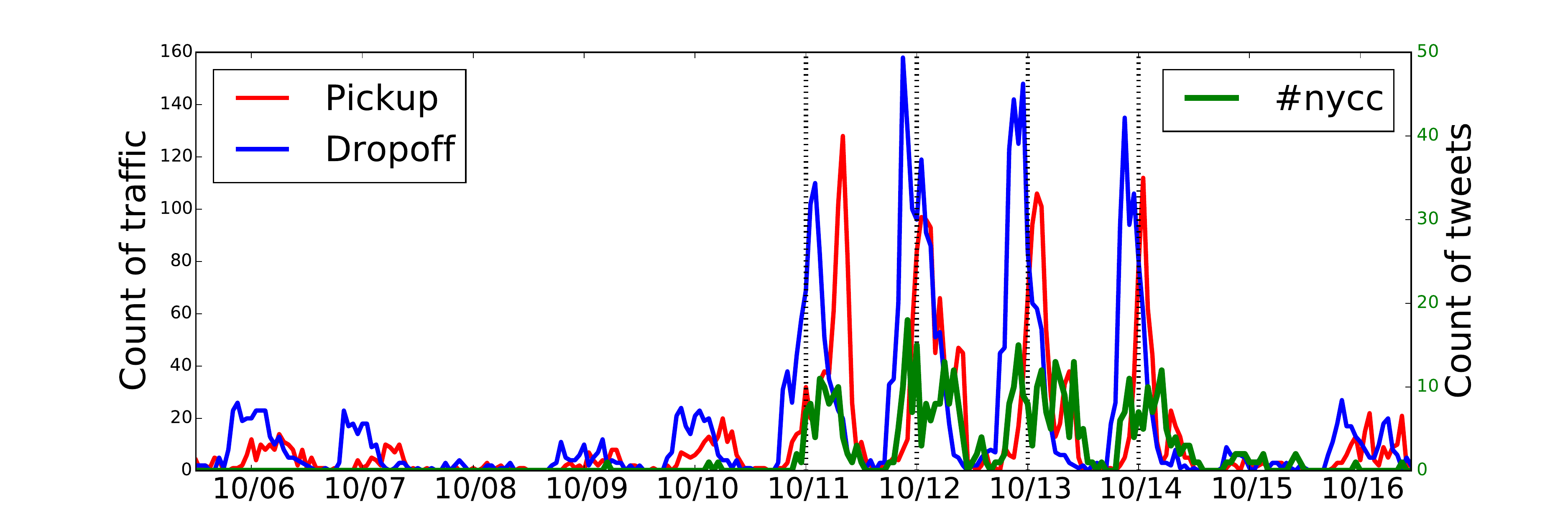}
  \caption{Correlation is observed between traffic and the frequency of tweets containing hashtag ``\#nycc'' (i.e., NYC comic con event).}
  ~\label{fig:nycc}
\end{figure}

\begin{figure}[t!]
\centering
  \includegraphics[width=0.7\columnwidth]{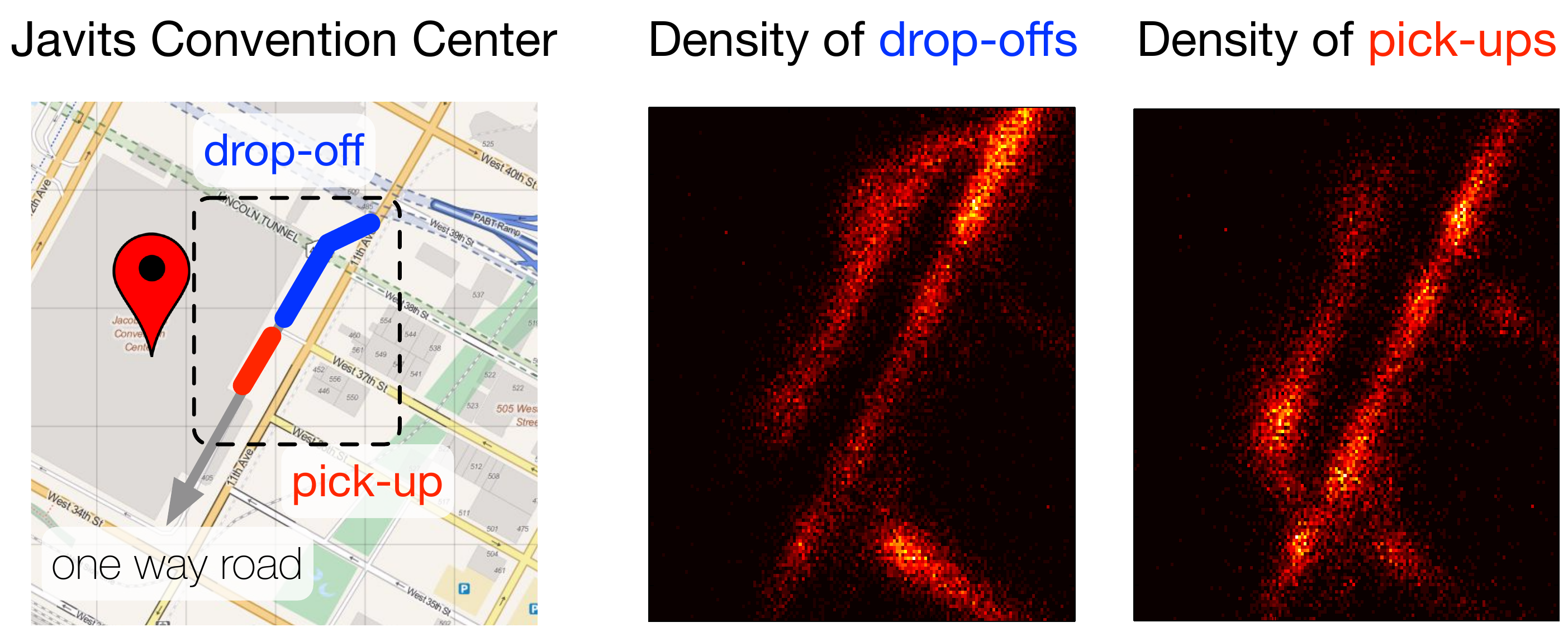}
  \caption{Traffic density near Javits Convention Center on an event day (October 12, 2012). High volume of drop-offs and pick-ups on the road of convention center entrance is observed. }
  ~\label{fig:jacob-traffic}
\end{figure}

To verify the correlation, we look at time series of taxi traffic with the frequency of hastag ``\#nycc''. As shown in Figure~\ref{fig:nycc}, there is a strong correlation among these three time series. In Figure~\ref{fig:jacob-traffic}, we further observe that the drop-offs usually occur before the front entrance and pick-ups occurs after the front entrance.

\subsubsection{Traffic w.r.t. Weather.} 
Intuitively, the extreme weather condition should impact the traffic. 
Here we particularly look at one weather event occurred during which time is covered by our datasets.
Hurricane Sandy is a category-3 major hurricane that hit New York City on Oct. 29, 2012. The wind speed attribute in our weather feature captures the signal of this disaster. As shown in Figure~\ref{fig:cases}(e), incorporating the weather information can effectively capture the significant drop of traffic volume during the time. 
By modelling the recovery time after the disaster, the fitted traffic pattern shows a slow recovery of traffic in the next 3 days, which aligns with the actual traffic pattern. Compared with not using the weather features, it demonstrates the utility of weather data in the presence of extreme weather conditions. 

\begin{figure}[t]
\centering
  \includegraphics[width=0.75\textwidth]{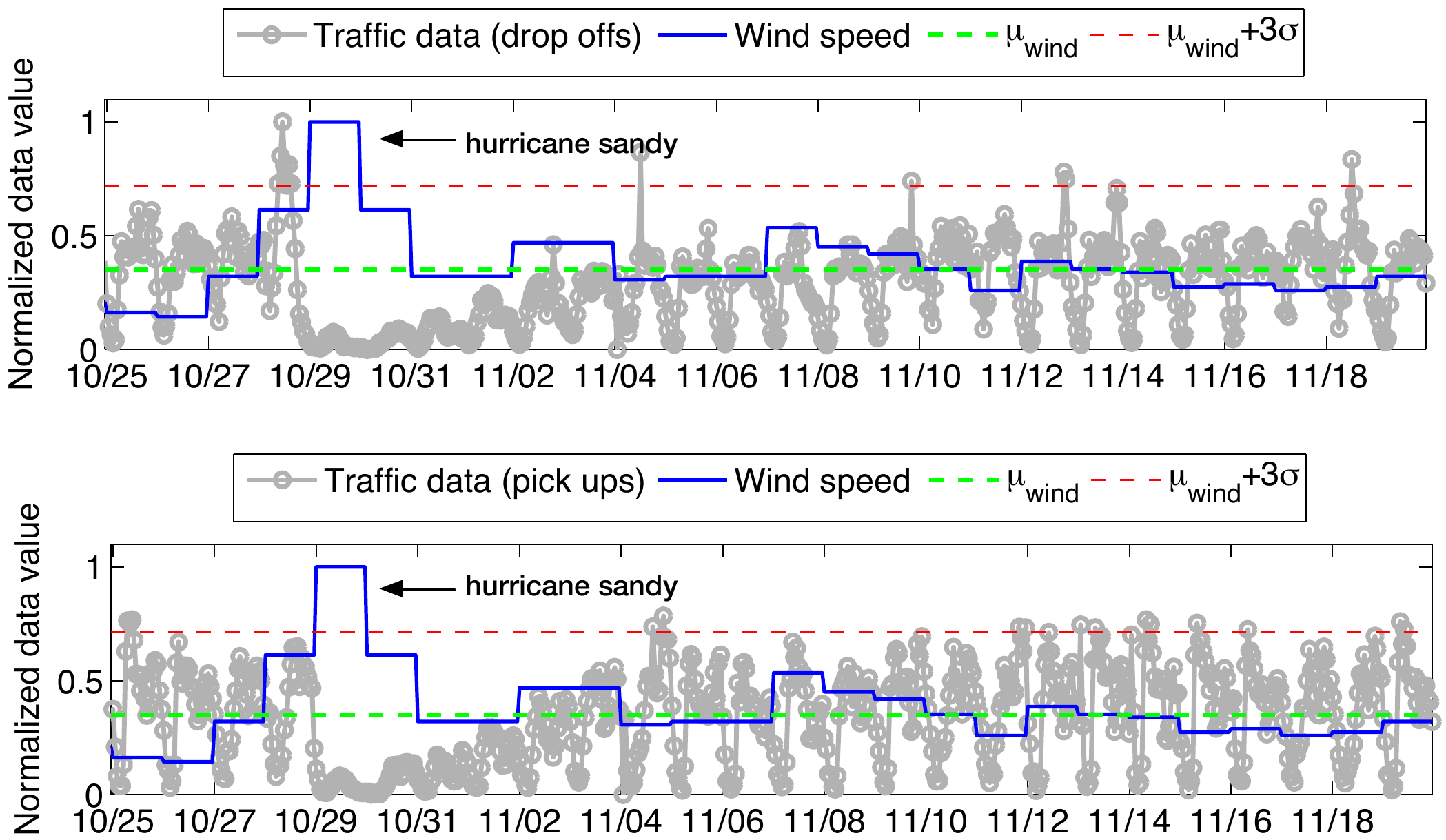}
  \caption{Traffic and weather. Hurricane Sandy arrived at NYC on 10/29/2012. The fastest 2-minute wind speed (WSF2) on that day is 17 meter per second. The average WSF2 is 6.2m/s for year 2012. }
  ~\label{fig:weather}
\end{figure}

\subsubsection{Insignificant Features.} 
While we found cases where many features have utilities in explain the traffic patterns, there are also features that have little impact on the traffic. We discuss two such features in this section, namely, the collisions and the snow fall.

\begin{figure}[h!]
\centering
  \includegraphics[width=0.75\textwidth]{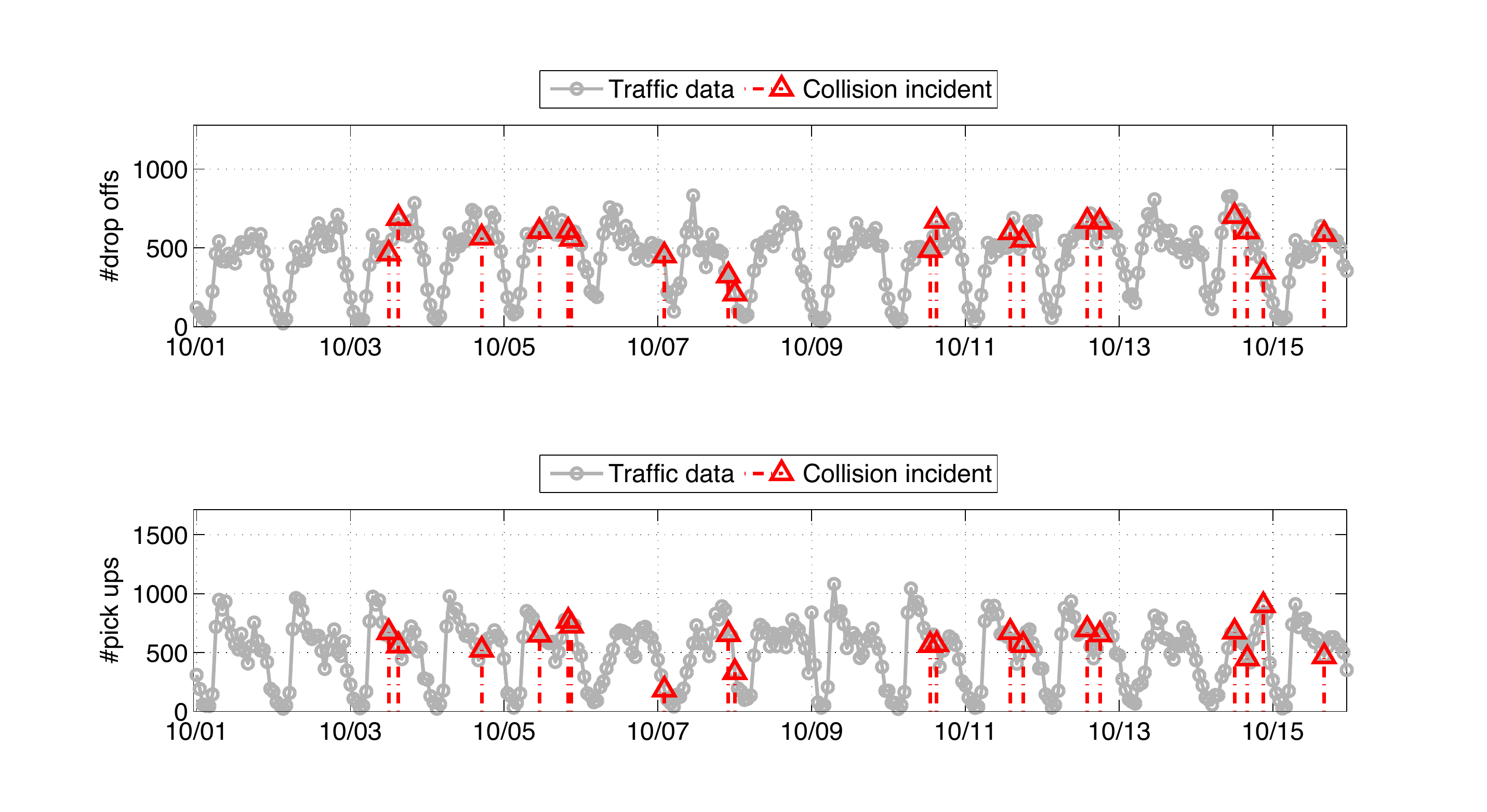}
  \caption{Collisions do not significantly impact traffic.}
  ~\label{fig:collision}
\end{figure}

\begin{figure}[h!]
\centering
  \includegraphics[width=0.75\textwidth]{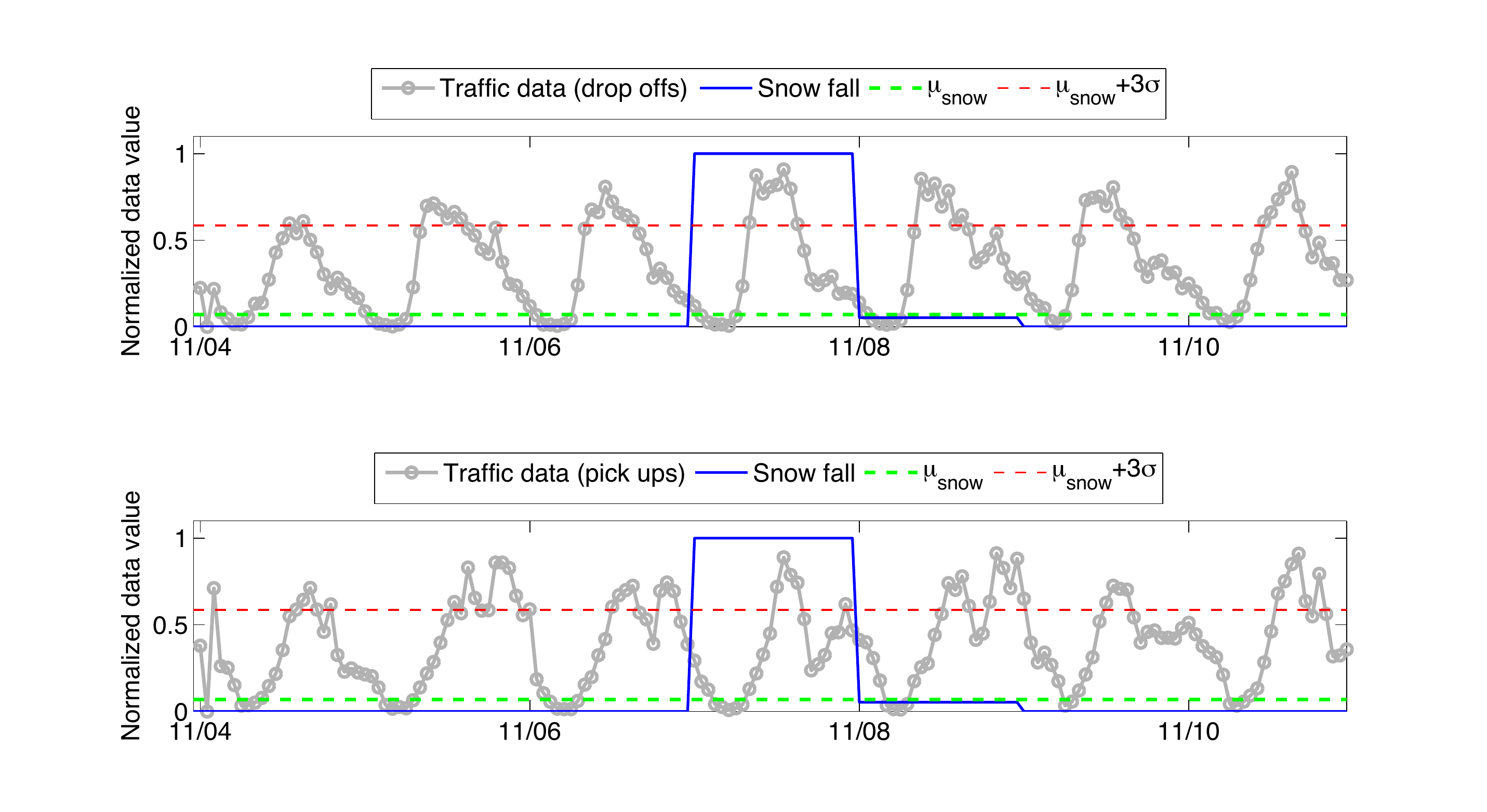}
  \caption{Snow storm on 11/07/2012 has a minor impact on traffic.}
  ~\label{fig:snowstorm}
\end{figure}

Common sense tells us that vehicle collisions cause blocking of road segments or traffic jam. Hence, there should be less pick-ups and drop-offs near the collision locations. However, we fail to see such a phenomenon in the taxi data. In Figure~\ref{fig:collision}, we show one month of traffic at Madison Square Garden together with the vehicle collisions in that region. We do not observe any significant drop in traffic volume near the time of the collisions. The reason could be that, while the collisions do affect the local traffic, its impact is very subtle on the overall traffic (in terms of number of pick-ups and drop-offs). Capturing such minor impact using urban data still remains a challenging problem.

Meanwhile, from November 7 to November 10, 2012, a nor'easter brought significant snow to the Northeastern United States. Before the nor'easter struck, officials recommended residents in low-lying areas of New York City to evacuate. Airlines canceled over 1,300 flights in and out of New York airports. Parks in New York City were closed, and construction was halted. However, even though a high volume of snowfall was reported on Nov. 7th, as shown in Figure~\ref{fig:snowstorm}, the traffic volume exhibits similar patterns as normal days except that there is a slight drop in the afternoon on Nov. 7th.
It is not certain whether such small variation in the traffic pattern is a result of the weather condition or due to the normal fluctuations of traffic.


\subsection{Application: Regular Traffic Forecasting}
\begin{figure}[t!]
\centering

  \includegraphics[width=0.9\textwidth]{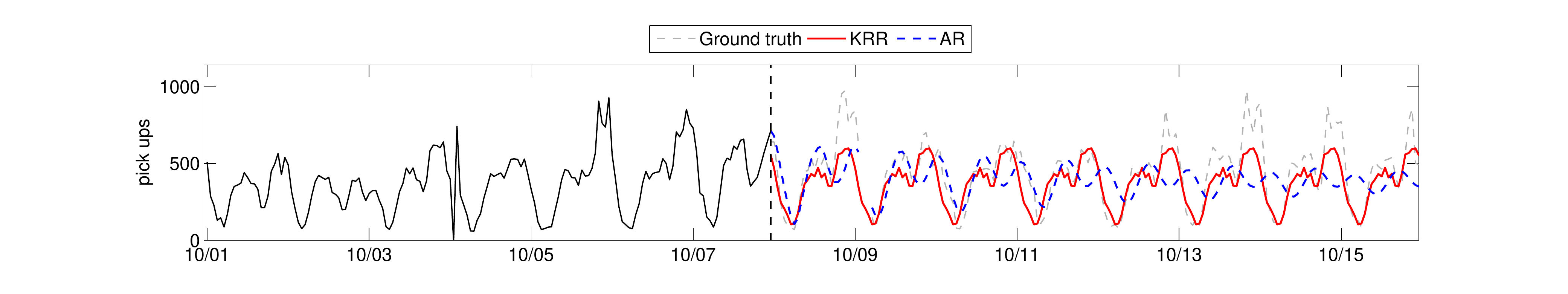} \\
  \includegraphics[width=0.9\textwidth]{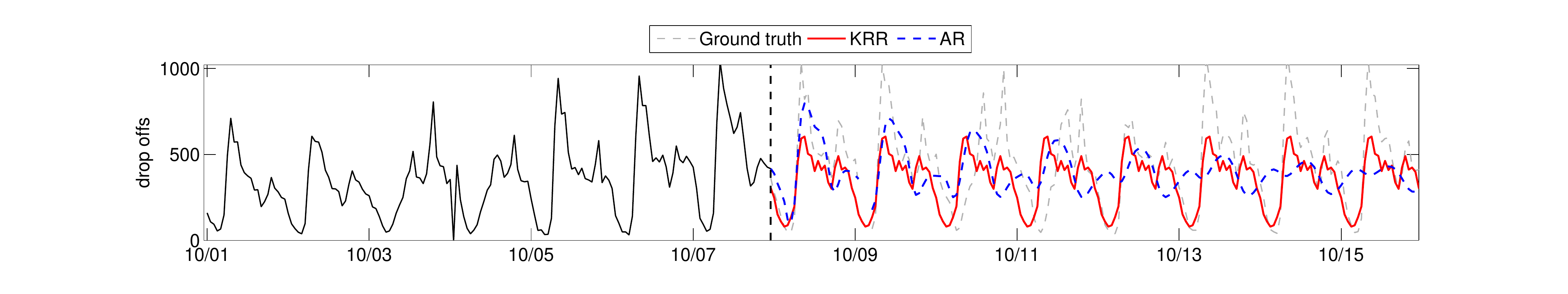}  
  \caption{Forecasting result on Times Square. The models are trained using data of the first week in October 2012 (\ie, solid black line). Forecasting starts in the following week (at the vertical dotted line). }
\label{fig:prediction}
\end{figure}

A better modelling of traffic with contextual urban data can be used for traffic forecasting.
In this section, we compare our approach with a forecasting method, autoregressive moving-average model (ARMA), that aims to capture the dynamics of traffic data with its own historical data.
ARMA is commonly applied for time series fitting and forecasting. 
The model has two parameters $p$ and $q$ for its auto regressive and moving average component, respectively. 
Both parameters specify the range of previous values that a prediction depends on.  
In our setting, we set $p=24$ hours and $q=1$.
As urban data reflecting irregular events are hard to known beforehand, we only use POI features for the prediction and aim to predict the regular part of the traffic patterns.

Our method achieves better forecasting result, \ie, ARMA gives negative $R^2$ result, which indicates a very bad prediction result. Our approach consistently gives $R^2$ around $0.5$ for all locations. 
We further investigate the performance of both approaches for forecasting regular traffic.
Figure~\ref{fig:prediction} shows an example of forecasting result at Times Squares.
We can see that the prediction of ARMA starts to converge and the peaks of the curve shift away from the actual data.
ARMA performs poorly, as it is known to deteriorates when used to make predictions for a time point further into the future.

\section{Discussion}
\label{sec:discussion}
\subsection{Key Findings}
Our results show that ubiquitous urban data can effectively explain the city traffic. Specifically, we have the observations:
\begin{itemize}[leftmargin=*]
\parskip -0.5ex

\item POI data can well describe the regular traffic patterns in a local region. 
The result is shown both quantitatively and qualitatively.
 
\item Geo-tagged tweets may be related to traffic dynamics caused by the local events. However, such a correlation is more obvious in relatively isolated locations (e.g., Javits Convention Center, which locates on the west boundary of Manhattan). 
 
\item Weather data are useful to describe natural disaster events. In particular, we observe hurricane Sandy from wind speed data and at the same time observe a significant drop of traffic volume in those days. However, not all the weather conditions have the same impacts. For example, the snow storm in November 2012 has a much smaller impact on the traffic compared to Sandy.

\item Vehicle collisions have little impact on hourly traffic. The reason could be (1) most collisions are minor accidents and do not have noticeable impact on traffic; (2) many major collisions happen on the highway or during the midnight, thus may not affect taxi pick-up and drop-off; (3) the temporal resolution we are currently using is hour but the collision impact may only last for a few minutes.
\item For insignificant changes in traffic, it is hard to tell whether the change is due to external impacts or to the normal fluctuations.
\end{itemize}

\subsection{Limitations of Current Method}

While most existing studies focus only on traffic data, our goal is to utilize ubiquitous urban datasets to interpret the traffic dynamics. But our finding may contain bias due to several limitations:
\begin{itemize}[leftmargin=*]
\parskip -0.5ex

\item The traffic at one location may not be well explained by the external factors at the \emph{same} location. There could be a cascaded effect of events at nearby locations. For example, if there is an accident on a major road in the morning peak hour, it may cause traffic jam on all in-bound routes. So a traffic jam does not mean that there is an accident at that particular location. In our currently analysis, we did not model such cascading effect.

\item We only consider traffic reflected by the numbers of drop-offs and pick-ups.
Natural, the traffic can be expressed in other forms, such as the travel time and the speed of taxi trips. The correlation between some external factors with the traffic more become more obvious. For example, collisions could potentially slow down the traffic but not necessarily impact the volume.

\item Taxi traffic is only a subset of entire traffic in a city. Considering additional traffic data such as public transportation and private vehicles should provide us a more complete view. At the same time, the external context data
may contain bias~\cite{HeSm14} as well, \eg, check-in data better reflect the behaviors of urban population rather than rural population.
It is an interesting future direction to consider more comprehensive datasets.

\item Correlations do not necessarily indicate causality. For example, an event appears to correlate with the traffic, but people might be just going to other events at the same location. We plan to pursue a causality inference model, which provides us with more confident and more fine-grained interpretations.
\end{itemize}

\section{Conclusion}
\label{sec:conclusion}
We explored the potentials of using ubiquitous urban datasets to interpret traffic data. The traffic data we use are pick-up and drop-off of taxi trips in New York City. The explanatory urban datasets include POI data from FourSquare, geo-tagged tweets, weather, and vehicle collisions. We propose to use ridge regression with polynomial kernel to describe the non-linear non-additive relationships of impacting factors. Both quantitative and qualitative studies help us better understand the relation between traffic with its context. Additionally, we show an interesting application on traffic prediction using external urban datasets. Finally, we discuss the key insights of our results and limitations of current method. We conclude that using ubiquitous urban datasets helps us better understand the urban dynamics and could benefit a set of applications such as smart city and intelligent transportation system.

\bibliographystyle{ACM-Reference-Format-Journals}
\bibliography{ref_jessie}


\begin{thebibliography}{00}


\ifx \showCODEN    \undefined \def \showCODEN     #1{\unskip}     \fi
\ifx \showDOI      \undefined \def \showDOI       #1{{\tt DOI:}\penalty0{#1}\ }
  \fi
\ifx \showISBNx    \undefined \def \showISBNx     #1{\unskip}     \fi
\ifx \showISBNxiii \undefined \def \showISBNxiii  #1{\unskip}     \fi
\ifx \showISSN     \undefined \def \showISSN      #1{\unskip}     \fi
\ifx \showLCCN     \undefined \def \showLCCN      #1{\unskip}     \fi
\ifx \shownote     \undefined \def \shownote      #1{#1}          \fi
\ifx \showarticletitle \undefined \def \showarticletitle #1{#1}   \fi
\ifx \showURL      \undefined \def \showURL       #1{#1}          \fi

\bibitem[\protect\citeauthoryear{Abadi, Rajabioun, and Ioannou}{Abadi
  et~al\mbox{.}}{2015}]%
        {ARI15}
{Afshin Abadi}, {Tooraj Rajabioun}, {and} {Petros~A Ioannou}. 2015.
\newblock \showarticletitle{Traffic flow prediction for road transportation
  networks with limited traffic data}.
\newblock {\em Intelligent Transportation Systems, IEEE Transactions on\/}
  {16}, 2 (2015), 653--662.
\newblock


\bibitem[\protect\citeauthoryear{Center}{Center}{2016}]%
        {nhc}
{National~Hurricane Center}. 2016.
\newblock Saffir-Simpson Hurricane Wind Scale.
\newblock   (2016).
\newblock
\newblock
\shownote{\url{http://www.nhc.noaa.gov/aboutsshws.php}.}


\bibitem[\protect\citeauthoryear{Centers}{Centers}{2016}]%
        {snow}
{NOAA's~National Centers}. 2016.
\newblock The Northeast Snowfall Impact Scale (NESIS).
\newblock   (2016).
\newblock
\newblock
\shownote{\url{https://www.ncdc.noaa.gov/snow-and-ice/rsi/nesis}.}


\bibitem[\protect\citeauthoryear{Chan, Dillon, Singh, and Chang}{Chan
  et~al\mbox{.}}{2012}]%
        {CDSC12}
{Kit~Yan Chan}, {Tharam~S Dillon}, {Jaipal Singh}, {and} {Elizabeth Chang}.
  2012.
\newblock \showarticletitle{Neural-network-based models for short-term traffic
  flow forecasting using a hybrid exponential smoothing and
  Levenberg--Marquardt algorithm}.
\newblock {\em Intelligent Transportation Systems, IEEE Transactions on\/}
  {13}, 2 (2012), 644--654.
\newblock


\bibitem[\protect\citeauthoryear{Chen, Hu, Meng, and Zhang}{Chen
  et~al\mbox{.}}{2011}]%
        {CHMZ11}
{Chenyi Chen}, {Jianming Hu}, {Qiang Meng}, {and} {Yi Zhang}. 2011.
\newblock \showarticletitle{Short-time traffic flow prediction with ARIMA-GARCH
  model}. In {\em Intelligent Vehicles Symposium (IV), 2011 IEEE}. IEEE,
  607--612.
\newblock


\bibitem[\protect\citeauthoryear{Chen, Chen, and Qian}{Chen
  et~al\mbox{.}}{2014}]%
        {CCQ14}
{Po-Ta Chen}, {Feng Chen}, {and} {Zhen Qian}. 2014.
\newblock \showarticletitle{Road traffic congestion monitoring in social media
  with hinge-loss Markov random fields}. In {\em Data Mining (ICDM), 2014 IEEE
  International Conference on}. IEEE, 80--89.
\newblock


\bibitem[\protect\citeauthoryear{Department}{Department}{2016}]%
        {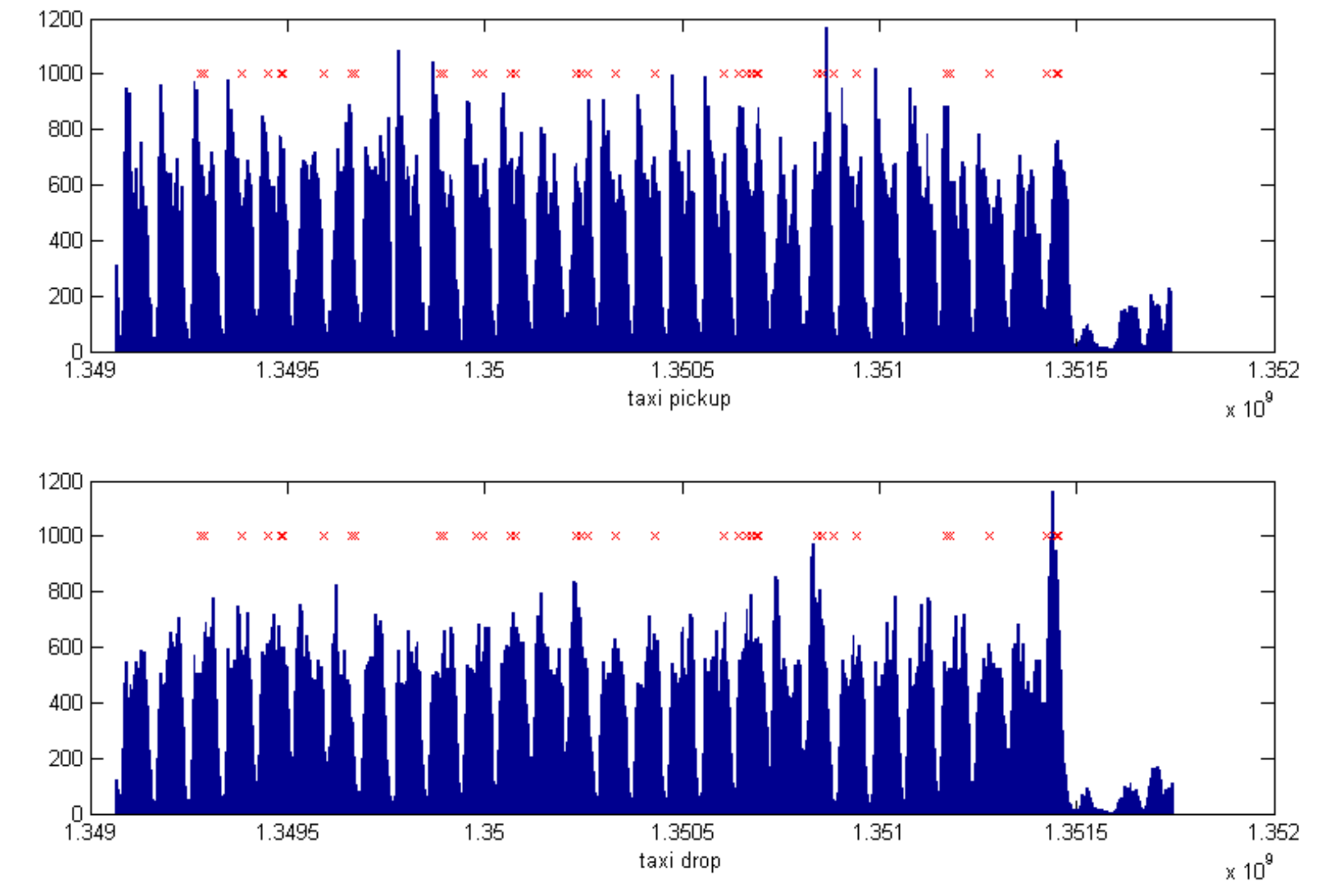}
{New York~Police Department}. 2016.
\newblock Motor Vehicle Collisions.
\newblock   (2016).
\newblock
\newblock
\shownote{\url{https://data.cityofnewyork.us/}.}


\bibitem[\protect\citeauthoryear{Fan, Song, Shibasaki, and Adachi}{Fan
  et~al\mbox{.}}{2015}]%
        {FSSA15}
{Zipei Fan}, {Xuan Song}, {Ryosuke Shibasaki}, {and} {Ryutaro Adachi}. 2015.
\newblock \showarticletitle{CityMomentum: an online approach for crowd behavior
  prediction at a citywide level}. In {\em Proceedings of the 2015 ACM
  International Joint Conference on Pervasive and Ubiquitous Computing}. ACM,
  559--569.
\newblock


\bibitem[\protect\citeauthoryear{for Environmental~Information}{for
  Environmental~Information}{2016}]%
        {noaa}
{NOAA's National~Centers for Environmental~Information}. 2016.
\newblock \url{http://www.ncdc.noaa.gov/}.
\newblock   (2016).
\newblock


\bibitem[\protect\citeauthoryear{Fourquare}{Fourquare}{2016}]%
        {foursquare}
{Fourquare}. 2016.
\newblock \url{https://foursquare.com/}.
\newblock   (2016).
\newblock


\bibitem[\protect\citeauthoryear{Government}{Government}{2016}]%
        {nyctaxi}
{NYC Government}. 2016.
\newblock NYC Taxi and Limousine Commission.
\newblock   (2016).
\newblock
\newblock
\shownote{\url{http://www.nyc.gov/html/tlc/html/about/trip_record_data.shtml}.}


\bibitem[\protect\citeauthoryear{Hecht and Stephens}{Hecht and
  Stephens}{2014}]%
        {HeSm14}
{Brent Hecht} {and} {Monica Stephens}. 2014.
\newblock \showarticletitle{A Tale of Cities: Urban Biases in Volunteered
  Geographic Information.}
\newblock {\em ICWSM\/}  {14} (2014), 197--205.
\newblock


\bibitem[\protect\citeauthoryear{Jeung, Shen, and Zhou}{Jeung
  et~al\mbox{.}}{2008}]%
        {JSZ08}
{H. Jeung}, {H.~T. Shen}, {and} {X. Zhou}. 2008.
\newblock \showarticletitle{Convoy Queries in Spatio-Temporal Databases}. In
  {\em Proc. 2008 Int. Conf. Data Engineering (ICDE'08)}. Cancun, Mexico,
  1457--1459.
\newblock


\bibitem[\protect\citeauthoryear{Lian, Ge, Zhang, Yuan, Xie, Zhou, and
  Rui}{Lian et~al\mbox{.}}{2015}]%
        {Lian+15}
{Defu Lian}, {Yong Ge}, {Fuzheng Zhang}, {Nicholas~Jing Yuan}, {Xing Xie}, {Tao
  Zhou}, {and} {Yong Rui}. 2015.
\newblock \showarticletitle{Content-Aware Collaborative Filtering for Location
  Recommendation Based on Human Mobility Data}. In {\em Data Mining (ICDM),
  2015 IEEE International Conference on}. IEEE, 261--270.
\newblock


\bibitem[\protect\citeauthoryear{Matsubara, Sakurai, and Faloutsos}{Matsubara
  et~al\mbox{.}}{2015}]%
        {Mats+15}
{Yasuko Matsubara}, {Yasushi Sakurai}, {and} {Christos Faloutsos}. 2015.
\newblock \showarticletitle{The web as a jungle: Non-linear dynamical systems
  for co-evolving online activities}. In {\em Proceedings of the 24th
  International Conference on World Wide Web}. ACM, 721--731.
\newblock


\bibitem[\protect\citeauthoryear{Matsubara, Sakurai, Van~Panhuis, and
  Faloutsos}{Matsubara et~al\mbox{.}}{2014}]%
        {Mats+14}
{Yasuko Matsubara}, {Yasushi Sakurai}, {Willem~G Van~Panhuis}, {and} {Christos
  Faloutsos}. 2014.
\newblock \showarticletitle{FUNNEL: automatic mining of spatially coevolving
  epidemics}. In {\em Proceedings of the 20th ACM SIGKDD international
  conference on Knowledge discovery and data mining}. ACM, 105--114.
\newblock


\bibitem[\protect\citeauthoryear{Monreale, Pinelli, Trasarti, and
  Giannotti}{Monreale et~al\mbox{.}}{2009}]%
        {MPTG09}
{Anna Monreale}, {Fabio Pinelli}, {Roberto Trasarti}, {and} {Fosca Giannotti}.
  2009.
\newblock \showarticletitle{Wherenext: a location predictor on trajectory
  pattern mining}. In {\em Proceedings of the 15th ACM SIGKDD international
  conference on Knowledge discovery and data mining}. ACM, 637--646.
\newblock


\bibitem[\protect\citeauthoryear{Noulas, Scellato, Mascolo, and Pontil}{Noulas
  et~al\mbox{.}}{2011}]%
        {NSMP11}
{Anastasios Noulas}, {Salvatore Scellato}, {Cecilia Mascolo}, {and}
  {Massimiliano Pontil}. 2011.
\newblock \showarticletitle{Exploiting Semantic Annotations for Clustering
  Geographic Areas and Users in Location-based Social Networks.}
\newblock {\em The social mobile web\/}  {11} (2011), 02.
\newblock


\bibitem[\protect\citeauthoryear{Okutani and Stephanedes}{Okutani and
  Stephanedes}{1984}]%
        {OkSt84}
{Iwao Okutani} {and} {Yorgos~J Stephanedes}. 1984.
\newblock \showarticletitle{Dynamic prediction of traffic volume through Kalman
  filtering theory}.
\newblock {\em Transportation Research Part B: Methodological\/} {18}, 1
  (1984), 1--11.
\newblock


\bibitem[\protect\citeauthoryear{Shang, Zheng, Tong, Chang, and Yu}{Shang
  et~al\mbox{.}}{2014}]%
        {Shan+14}
{Jingbo Shang}, {Yu Zheng}, {Wenzhu Tong}, {Eric Chang}, {and} {Yong Yu}. 2014.
\newblock \showarticletitle{Inferring gas consumption and pollution emission of
  vehicles throughout a city}. In {\em Proceedings of the 20th ACM SIGKDD
  international conference on Knowledge discovery and data mining}. ACM,
  1027--1036.
\newblock


\bibitem[\protect\citeauthoryear{Song, Qu, Blumm, and Barabasi}{Song
  et~al\mbox{.}}{2010}]%
        {SQBB10}
{C. Song}, {Z. Qu}, {N. Blumm}, {and} {A.~L. Barabasi}. 2010.
\newblock \showarticletitle{Limits of predictability in human mobility}. In
  {\em Science}. 1018--1021.
\newblock


\bibitem[\protect\citeauthoryear{Song, Zhang, Sekimoto, Horanont, Ueyama, and
  Shibasaki}{Song et~al\mbox{.}}{2013}]%
        {Song+13}
{Xuan Song}, {Quanshi Zhang}, {Yoshihide Sekimoto}, {Teerayut Horanont},
  {Satoshi Ueyama}, {and} {Ryosuke Shibasaki}. 2013.
\newblock \showarticletitle{Modeling and probabilistic reasoning of population
  evacuation during large-scale disaster}. In {\em Proceedings of the 19th ACM
  SIGKDD international conference on Knowledge discovery and data mining}. ACM,
  1231--1239.
\newblock


\bibitem[\protect\citeauthoryear{Song, Zhang, Sekimoto, and Shibasaki}{Song
  et~al\mbox{.}}{2014}]%
        {Song+14}
{Xuan Song}, {Quanshi Zhang}, {Yoshihide Sekimoto}, {and} {Ryosuke Shibasaki}.
  2014.
\newblock \showarticletitle{Prediction of human emergency behavior and their
  mobility following large-scale disaster}. In {\em Proceedings of the 20th ACM
  SIGKDD international conference on Knowledge discovery and data mining}. ACM,
  5--14.
\newblock


\bibitem[\protect\citeauthoryear{Sun, Zhang, and Yu}{Sun et~al\mbox{.}}{2006}]%
        {SZY06}
{Shiliang Sun}, {Changshui Zhang}, {and} {Guoqiang Yu}. 2006.
\newblock \showarticletitle{A Bayesian network approach to traffic flow
  forecasting}.
\newblock {\em Intelligent Transportation Systems, IEEE Transactions on\/} {7},
  1 (2006), 124--132.
\newblock


\bibitem[\protect\citeauthoryear{Van Der~Voort, Dougherty, and Watson}{Van
  Der~Voort et~al\mbox{.}}{1996}]%
        {VDW96}
{Mascha Van Der~Voort}, {Mark Dougherty}, {and} {Susan Watson}. 1996.
\newblock \showarticletitle{Combining Kohonen maps with ARIMA time series
  models to forecast traffic flow}.
\newblock {\em Transportation Research Part C: Emerging Technologies\/} {4}, 5
  (1996), 307--318.
\newblock


\bibitem[\protect\citeauthoryear{Wu, Wang, and Li}{Wu et~al\mbox{.}}{2016}]%
        {WWL16}
{Fei Wu}, {Hongjian Wang}, {and} {Zhenhui Li}. 2016.
\newblock \showarticletitle{Interpreting traffic dynamics using ubiquitous
  urban data}. In {\em Proceedings of the 24th ACM SIGSPATIAL International
  Conference on Advances in Geographic Information Systems}. ACM, 69.
\newblock


\bibitem[\protect\citeauthoryear{Xie, Zhang, and Ye}{Xie et~al\mbox{.}}{2007}]%
        {XZY07}
{Yuanchang Xie}, {Yunlong Zhang}, {and} {Zhirui Ye}. 2007.
\newblock \showarticletitle{Short-Term Traffic Volume Forecasting Using Kalman
  Filter with Discrete Wavelet Decomposition}.
\newblock {\em Computer-Aided Civil and Infrastructure Engineering\/} {22}, 5
  (2007), 326--334.
\newblock


\bibitem[\protect\citeauthoryear{Xu, Kong, Klette, and Liu}{Xu
  et~al\mbox{.}}{2014}]%
        {XKKL14}
{Yanyan Xu}, {Qing-Jie Kong}, {Reinhard Klette}, {and} {Yuncai Liu}. 2014.
\newblock \showarticletitle{Accurate and interpretable Bayesian MARS for
  traffic flow prediction}.
\newblock {\em Intelligent Transportation Systems, IEEE Transactions on\/}
  {15}, 6 (2014), 2457--2469.
\newblock


\bibitem[\protect\citeauthoryear{Yuan, Zheng, and Xie}{Yuan
  et~al\mbox{.}}{2012}]%
        {YZX12}
{Jing Yuan}, {Yu Zheng}, {and} {Xing Xie}. 2012.
\newblock \showarticletitle{Discovering regions of different functions in a
  city using human mobility and POIs}. In {\em Proceedings of the 18th ACM
  SIGKDD international conference on Knowledge discovery and data mining}. ACM,
  186--194.
\newblock


\bibitem[\protect\citeauthoryear{Zhang and Wang}{Zhang and Wang}{2015}]%
        {ZhWa15}
{Wei Zhang} {and} {Jianyong Wang}. 2015.
\newblock \showarticletitle{Location and Time Aware Social Collaborative
  Retrieval for New Successive Point-of-Interest Recommendation}. In {\em
  Proceedings of the 24th ACM International on Conference on Information and
  Knowledge Management}. ACM, 1221--1230.
\newblock


\bibitem[\protect\citeauthoryear{Zheng and Ni}{Zheng and Ni}{2012}]%
        {ZhNi12}
{Jiangchuan Zheng} {and} {Lionel~M Ni}. 2012.
\newblock \showarticletitle{An unsupervised framework for sensing individual
  and cluster behavior patterns from human mobile data}. In {\em Proceedings of
  the 2012 ACM Conference on Ubiquitous Computing}. ACM, 153--162.
\newblock


\bibitem[\protect\citeauthoryear{Zheng, Capra, Wolfson, and Yang}{Zheng
  et~al\mbox{.}}{2014}]%
        {ZCWY14}
{Yu Zheng}, {Licia Capra}, {Ouri Wolfson}, {and} {Hai Yang}. 2014.
\newblock \showarticletitle{Urban computing: concepts, methodologies, and
  applications}.
\newblock {\em TIST\/} {5}, 3 (2014), 38.
\newblock


\bibitem[\protect\citeauthoryear{Zheng, Liu, and Hsieh}{Zheng
  et~al\mbox{.}}{2013}]%
        {ZLH13}
{Yu Zheng}, {Furui Liu}, {and} {Hsun-Ping Hsieh}. 2013.
\newblock \showarticletitle{U-Air: When urban air quality inference meets big
  data}. In {\em Proceedings of the 19th ACM SIGKDD international conference on
  Knowledge discovery and data mining}. ACM, 1436--1444.
\newblock


\bibitem[\protect\citeauthoryear{Zheng, Yi, Li, Li, Shan, Chang, and Li}{Zheng
  et~al\mbox{.}}{2015a}]%
        {Zhen+15}
{Yu Zheng}, {Xiuwen Yi}, {Ming Li}, {Ruiyuan Li}, {Zhangqing Shan}, {Eric
  Chang}, {and} {Tianrui Li}. 2015a.
\newblock \showarticletitle{Forecasting Fine-Grained Air Quality Based on Big
  Data}. In {\em Proc. ACM SIGKDD}.
\newblock


\bibitem[\protect\citeauthoryear{Zheng, Zhang, and Yu}{Zheng
  et~al\mbox{.}}{2015b}]%
        {ZZY15}
{Yu Zheng}, {Huichu Zhang}, {and} {Yong Yu}. 2015b.
\newblock \showarticletitle{Detecting collective anomalies from multiple
  spatio-temporal datasets across different domains}.
\newblock {\em Proc. 2015 ACM Int. Symp. Advances in Geographic Information
  Systems (GIS'15)\/} (2015).
\newblock


\end{thebibliography}
\end{document}